\documentclass[fleqn,usenatbib,usedcolumn]{mnras}

\usepackage[british]{babel}             
\usepackage{newtxtext}                  
   \usepackage[slantedGreek]{newtxmath}    
   \usepackage[T1]{fontenc}                
   \usepackage{graphicx}                   
\usepackage{lscape}
\usepackage{longtable}
\usepackage{tabu}
\usepackage{url}
\usepackage{breakurl}
\usepackage{bigints}
\renewcommand{\vec}[1]{\boldsymbol{#1}}

\title[Physical modelling with Einasto profile]{Physical modelling of galaxy cluster {Sunyaev--Zel'dovich data} using Einasto dark matter profiles}

\author[K. Javid et al.]
{Kamran Javid$^{1,2}$\thanks{E-mail: kj316@mrao.cam.ac.uk},
Yvette C. Perrott$^{1,3}$,
Clare Rumsey$^{1}$,
\newauthor
and Richard D. E. Saunders$^{1,2}$
\\
$^{1}$Astrophysics Group, Cavendish Laboratory, JJ Thomson Avenue, Cambridge CB3 0HE, UK\\
$^{2}$Kavli Institute for Cosmology Cambridge, Madingley Road, Cambridge, CB3 0HA, UK \\
$^{3}$School of Chemical and Physical Sciences, Victoria University of Wellington, PO Box 600, Wellington 6140, New Zealand\\
}

\date{Accepted XXX. Received YYY; in original form ZZZ}

\pubyear{2018}

\begin{document}
\label{firstpage}
\pagerange{\pageref{firstpage}--\pageref{lastpage}}
\maketitle

\begin{abstract}
We derive a model for Sunyaev--Zel'dovich data from a galaxy cluster which uses an Einasto profile to model the cluster's dark matter component. This model is similar to the physical models for clusters previously used by the Arcminute Microkelvin Imager (AMI) consortium, which model the dark matter using a Navarro-Frenk-White (NFW) profile, but the Einasto profile provides an extra degree of freedom. We thus present a comparison between two physical models which differ only in the way they model dark matter: one which uses an NFW profile (PM I) and one that uses an Einasto profile (PM II). 
We illustrate the differences between the models by plotting physical properties of clusters as a function of cluster radius. 
We generate AMI simulations of clusters which are \textit{created} and \textit{analysed} with both models. From this we find that for 14 of the 16 simulations, the Bayesian evidence gives no  preference to either of the models according to the Jeffreys scale, and for the other two simulations, weak preference in favour of the correct model. However, for the mass estimates obtained from the analyses, the values were within $1\sigma$ of the input values for 14 out of 16 of the clusters when using the correct model, but only in 6 out of 16 cases when the incorrect model was used to analyse the data.
Finally we apply the models to real data from cluster A611 obtained with AMI, and find the mass estimates to be consistent with one another except in the case of when PM II is applied using an extreme value for the Einasto shape parameter. 

\end{abstract}



\section{Introduction}
Clusters of galaxies are the most massive gravitationally bound objects known in the Universe, and as such sample the Universe's matter content. Some 85 to 90 percent (see e.g. \citealt{2006ApJ...640..691V}, \citealt{2009ApJ...692.1060V}, \citealt{2011ApJS..192...18K}, \citealt{2019A&A...621A..40E} and references in these for data and issues) of cluster total mass is in (non-baryonic) dark matter. Stars, gas and dust in galaxies, as well as a hot ionised intra-cluster medium (ICM) make up the remaining mass, with the latter being much the most massive baryonic component. The galaxies emit in the optical and infrared wavebands, the ICM emits in X-ray via thermal Bremsstrahlung, and interacts with cosmic microwave background (CMB) photons via inverse Compton scattering. This last effect is known as the Sunyaev--Zel'dovich (SZ) effect \citep{1970CoASP...2...66S}.

It is this effect that the physical modelling of clusters described in \citet{2012MNRAS.423.1534O} (from here on referred to as MO12) and \citet{2013MNRAS.430.1344O} aims to predict; this physical modelling has been applied in for example \citet{2018arXiv180501968J} (from here on KJ18) and \citet{2019MNRAS.483.3529J} (KJ19). The model presented in MO12 uses a Navarro-Frenk-White (NFW) profile \citep{1995MNRAS.275..720N} for the dark matter component of the galaxy cluster, which is derived from N-body simulations of galaxy clusters. \citet{1965TrAlm...5...87E} derives an empirical profile for dark matter halos. Previous investigations comparing the two dark matter profiles using simulated data (see e.g. \citealt{2014MNRAS.441.3359D}, \citealt{2014ApJ...797...34M}, \citealt{2016MNRAS.457.4340K} and \citealt{2016JCAP...01..042S}) have shown that the Einasto model provides a better fit. In particular, \citet{2016JCAP...01..042S} showed that, for weak lensing analysis of clusters, the NFW profile can overestimate virial masses of very massive halos ($\geq 10^{15}M_{\mathrm{Sun}} / h$ where $M_{\mathrm{Sun}}$ is units of solar mass and $h \equiv h_{100} = H_{0} / (100~\mathrm{km}~\mathrm{s}^{-1}~\mathrm{Mpc}^{-1})$ where $H_{0}$ is the value of the Hubble constant now) by up to 10\%. Errors of this magnitude are non-negligible and can have substantial effects on the estimates of parameters such as the normalisation of the matter density fluctuations $\sigma_{8}$, the matter density $\Omega_{\rm M}$, and the density and equation of state of the dark energy field parameters $\Omega_{\mathrm{DE}}$ and $w$.
 
It is these previous analyses which have motivated us to derive a physical galaxy cluster model for interferometric SZ data which uses the Einasto profile to model the dark matter component of the cluster. For a range of cluster model inputs we compare the physical parameter profiles of the two models, to see how the clusters they represent deviate from one another. We then compare the parameter estimates and fits of the NFW and Einasto models for simulated cluster data created with both Einasto and NFW profiles (see e.g. \citealt{2002MNRAS.333..318G}), as well as real data for cluster A611 obtained from the Arcminute Microkelvin Imager (AMI) radio interferometer system \citep{2008MNRAS.391.1545Z}, \citep{2018MNRAS.475.5677H}. This is to see how flexible the models are in modelling clusters generated with different profiles, and test how well they fit real data.

Section~\ref{sec:interferomSZ} of this paper gives a brief overview of the theory behind interferometry and the SZ effect. In Section~\ref{sec:model} we derive the physical model for interferometric SZ data which uses the Einasto profile, and describe the Bayesian methodology for fitting a model to the data. Section~\ref{sec:results} present the results of our analysis, including the radial profiles of physical parameters using both the model presented here and the one derived in MO12 for a range of clusters. We also present the results of applying these models to simulated and real cluster data using Bayesian analysis.

A `concordance' flat $\Lambda$CDM cosmology is assumed: $\Omega_{\rm M} = 0.3$, $\Omega_{\Lambda} = 0.7$, $\Omega_{\rm R} = 0$, $\Omega_{\rm K} = 0$, $\sigma_{8} = 0.8$, $w_{0} = -1$, and $w_{\rm a} = 0$. The first four parameters correspond to the (dark + baryonic) matter, the cosmological constant, the radiation, and the curvature densities respectively. $\sigma_{8}$ is the power spectrum normalisation on the scale of $8$~$h^{-1}$~Mpc now.
$w_{\rm 0}$ and $w_{\rm a}$ are the equation of state parameters of the Chevallier-Polarski-Linder parameterisation \citep{2001IJMPD..10..213C}.


\section{Measuring the SZ effect with an Interferometer}
\label{sec:interferomSZ}
For a small field size, an interferometer samples from the two-dimensional complex visibility plane $\vec{u}$, also known as the $u$-$v$ plane. At frequency $\nu$, the samples correspond to the Fourier components of the sky brightness distribution $\tilde{I}_{\nu}(\vec{u})$. $\tilde{I}_{\nu}(\vec{u})$ is given by the weighted Fourier transform of the surface brightness $I_{\nu}(\vec{x})$
\begin{equation}\label{eqn:surfbrightFT}
\tilde{I}_{\nu}(\vec{u}) = \int_{-\infty}^{\infty} A_{\nu}(\vec{x})I_{\nu}(\vec{x})e^{2\pi i\vec{u}\cdot\vec{x}}\,\mathrm{d}^{2}\vec{x},
\end{equation}
where $\vec{x}$ is the position in the sky relative to the phase centre and $A_{\nu}(\vec{x})$ is the primary beam of the antennas for a given frequency. The positions at which $\tilde{I}_{\nu}(\vec{u})$ are sampled from is therefore determined by the physical orientation of the antennas. 
The change in CMB intensity $\delta I$, due to the thermal SZ effect in a galaxy cluster is given by (see e.g. \citealt{1999PhR...310...97B})
\begin{equation}\label{eqn:surfbright}
\delta I_{\rm{cl},\nu} = T_{\rm CMB}yf_{\nu}\frac{\partial B_{\nu}(T)}{\partial T}\bigg|_{ T = T_{\rm CMB}},
\end{equation}
where the last factor is the derivative of the blackbody spectrum with respect to temperature evaluated at the absolute temperature of the CMB, which at present is $T_{\rm CMB} = 2.728~\mathrm{K}$ \citep{1996ApJ...473..576F}. $B_{\nu}(T)$ is the spectral radiance of blackbody radiation (given by Planck's law).
The function $f_{\nu}$ expresses the spectral dependence of the SZ signal and is derived from the Kompaneets equation \citep{1957JETP...4..730}. 

\citet{1995ARA&A..33..541R} states that for the unmodified Kompaneets equation to be valid, the optical depth of the cluster $\tau$, must be sufficiently large to justify using a diffusion approximation for the scattering process. It is clear that at AMI observing frequencies $h_{\rm P}\nu \ll m_{\rm e}c^{2}$ where $h_{\rm P}$ is the Planck constant,  $m_{\rm e}$ is the mass of an electron and $c$ is the speed of light, the photons scatter in the Thomson limit. In this limit the scattering rate is $\propto \sigma_{\rm T}n_{\rm e}$ where $\sigma_{\rm T}$ is the Thomson scattering cross-section and $n_{e}$ is the electron number density in the ICM. Thus the optical depth is given by
\begin{equation}\label{eqn:optdepth}
\tau = \int n_{\rm e}(r)\sigma_{\rm T} \mathrm{d}l,
\end{equation}
where $r$ is the radius from the galaxy cluster centre and the integral is along the line of sight. The non-relativistic form for $f_{\nu}$ is given by
\begin{equation}\label{eqn:funcofnu}
f_{\nu} = X\coth(X/2) - 4,
\end{equation}
where
\begin{equation}\label{eqn:nufuncfact}
X = \frac{h_{\rm P}\nu}{k_{\rm B}T_{\rm CMB}}.
\end{equation}
Here $k_{\rm B}$ is the Boltzmann constant. 
Referring back to equation~\ref{eqn:surfbright}, $y$ is the Comptonisation parameter which is the number of collisions multiplied by the mean fractional change in energy of the photons per collision, integrated along the line of sight. On average the electrons in the ICM transfer an energy $k_{\rm B} T_{\rm e}(r) / m_{\rm e}c^{2}$ to the scattered CMB photons, where $T_{\rm e}(r)$. In the Thomson scattering regime described above this leads to
\begin{equation}\label{eqn:comptparam}
y = \frac{\sigma_{\rm T}k_{\rm B}}{m_{\rm e}c^{2}}\int T_{\rm e}(r) n_{\rm e}(r)\,\mathrm{d}l.
\end{equation}
If an ideal gas equation of state is assumed for the electron gas then in terms of the electron pressure $P_{\mathrm{e}}(r)$, the Comptonisation parameter is given by
\begin{equation}\label{eqn:comptparamideal}
y = \frac{\sigma_{\rm T}}{m_{\rm e}c^{2}}\int P_{\rm e}(r)\,\mathrm{d}l.
\end{equation}
Combining equations \ref{eqn:surfbright}, \ref{eqn:funcofnu} and \ref{eqn:comptparamideal}, we arrive at the following expression for $\delta I_{\nu , \mathrm{cl}}$ in the non-relativistic limit
\begin{equation}\label{eqn:surfbrightcomplete}
\delta I_{\mathrm{cl}, \nu } = \frac{2\sigma_{\rm T}(k_{\rm B}T_{\rm CMB})^{3} X^{4}e^{X}}{h_{\rm P}^2c^{4}(e^{X}-1)^{2}}[X\coth(X/2) - 4]\int P_{\rm e}(r)\,\mathrm{d}l.
\end{equation}

Relativistic treatments of $f_{\nu}$ have been considered in e.g. \citet{1995ARA&A..33..541R}, \citet{1998ApJ...502....7I}, \citet{1998ApJ...499....1C}, \citet{1998ApJ...508...17N}, \citet{1998A&A...336...44P}, and more recently \citet{2012MNRAS.426..510C}, by incorporating relativistic terms into the Kompaneets equation. Relativistic effects may be important in clusters where the ICM temperatures are high. Indeed \citet{1994ApJ...436L..67A} and \citet{1996ApJ...456..437M} have shown that electrons in the ICM can reach energies above $10$~keV. Challinor and Lasenby show that these effects lead to a small decrease in the SZ effect. 
We have calculated the relativistic correction at 15.5\,GHz using the \textsc{SZPack} code of \citet{2012MNRAS.426..510C} and find that to a good approximation $(\delta I_{\mathrm{cl,rel}}-\delta I_{\mathrm{cl,non-rel}}) / \delta I_{\mathrm{cl,rel}} = -0.0034 T_{\rm e}$, where $T_{\rm e}$ is the temperature of ICM electrons, with the correction only reaching 5\% at $\approx$\,15\,keV. This is sub-dominant to other forms of error for AMI data and we neglect it for the current analysis, although it may become important for future instruments. 


\section{Modelling interferometric SZ data}
\label{sec:model}

We first discuss how to model $\delta I_{\nu , \mathrm{cl}}$ arising from the SZ effect, starting from input parameters (see Section~\ref{subsec:bayesian} for more on what input parameters are) which describe some physical properties of a cluster. Equation \ref{eqn:surfbrightFT} can then be used to replicate the quantity measured by an interferometer. \\
In Section~\ref{subsec:bayesian} we discuss how to use Bayesian inference to perform parameter estimation and model selection for comparison between different models using the NFW and Einasto dark matter profiles. 


\subsection{Cluster models}
\label{subsec:clustermodels}

We now consider two cases for modelling physical properties of galaxy clusters, which we denote PM I and PM II. PM I is as described in MO12, but with the alteration in the mapping from $r_{200}$ to $r_{500}$ (defined below) described in KJ19. PM~II is qualitatively similar to PM~I, but with an Einasto profile replacing the NFW profile used for the dark matter component in PM~I.

Three cluster input parameters are required for either PM: $M(r_{200})$ which is the mass enclosed up to radius $r_{200}$ from the cluster centre, $f_{\rm gas}(r_{200})$ which is the fraction of the total mass attributed to the gas mass enclosed up to radius $r_{200}$, and $z$, the redshift of the cluster. A fourth input parameter is required for the PM II which we call the Einasto parameter $\alpha_{\rm Ein}$, which is also described below. 
Note that in general the radius $r_{\upDelta}$ is the radius from the centre at which the average total enclosed mass density is $\upDelta$ times $\rho_{\rm crit}(z)$, the critical density at the $z$ of the cluster. $\rho_{\rm crit}(z)$ is given by $\rho_{\rm crit}(z) = 3H(z)^{2}/8\pi G$ where $H(z)$ is the Hubble parameter and $G$ is Newton's constant.
Note further that the total mass out to $r_{\upDelta}$ is given by 
\begin{equation}
\label{eqn:massrhocrit}
M(r_{\upDelta}) = \frac{4\pi}{3} \upDelta \rho_{\rm crit} (z) r_{\upDelta}^{3}. 
\end{equation}
Hence $r_{200}$ can be calculated from $M(r_{200})$. \\

We assume spherical symmetry, hydrostatic equilibrium, and that the cluster gas is an ideal gas.  Both models follow the same general computational steps, as follows.
\begin{enumerate}
\item{Given an input $M(r_{200})$, $z$ and $f_{\rm gas}(r_{200})$, the normalisation of the dark matter mass profile is fixed by the requirement that $M_{\rm dm}(r_{200}) = (1-f_{\rm gas}(r_{200})) M(r_{200})$ and the dark matter mass profile $M_{\rm dm}(r)$ is then fully specified.}
\item{We make the further assumption that we can calculate an initial estimate of the total mass profile of the cluster using only the dark matter model profile, i.e.\ making the approximation that all of the mass in the cluster is dark matter; or, equivalently, that the shape of the sum of the dark matter and gas mass profiles resembles the shape of the dark-matter-only profile
\begin{align}
\label{eqn:massapprox}
M(r) &= \int_{0}^{r} 4\pi r'^{2} \left ( \rho_{\rm dm}(r') + \rho_{\rm g}(r') \right ) \, \rm{d}r'\nonumber\\
 &\approx \int_{0}^{r} 4\pi r'^{2} \frac{\rho_{\rm dm}(r')}{1-f_{\rm gas}(r_{200})} \, \rm{d}r',
\end{align}
where $\rho_{\rm dm}(r)$ and $\rho_{\rm g}(r)$ are the dark matter and gas density profiles respectively.
}
\item{Now we can analytically solve the hydrostatic equilibrium equation $\frac{\mathrm{d}P_{\rm g}(r)}{\mathrm{d}r} = -\rho_{\rm g}(r)\frac{GM(r)}{r^{2}}$ for the gas mass density profile $\rho_{\rm g}$, given a template for the gas pressure $P_{\rm g}(r)$ and the initial dark-matter-only estimate for $M(r)$.}
\item{We then numerically integrate the $\rho_{\rm g}$ solution to get the gas mass profile $M_{\rm g}(r)$ and fix the gas mass normalisation at $r_{200}$ using the input $f_{\rm gas}(r_{200})$.}
\item{Finally, we derive the pressure profile normalisation from the gas mass normalisation and calculate the SZ signal.}
\end{enumerate}

\begin{figure*}
  \begin{center}
  \includegraphics[width=0.5\textwidth]{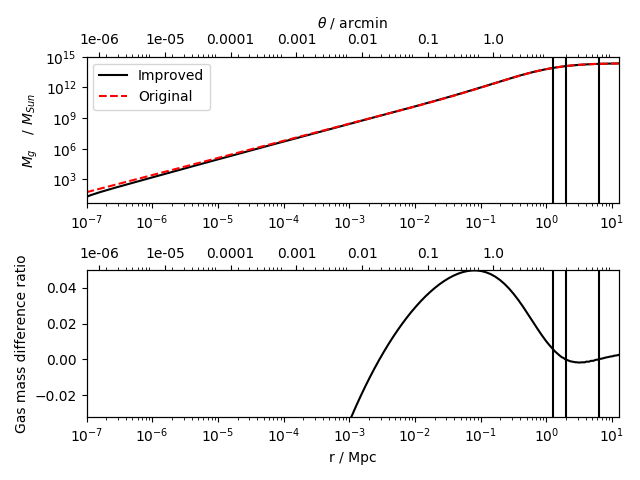}\includegraphics[width=0.5\textwidth]{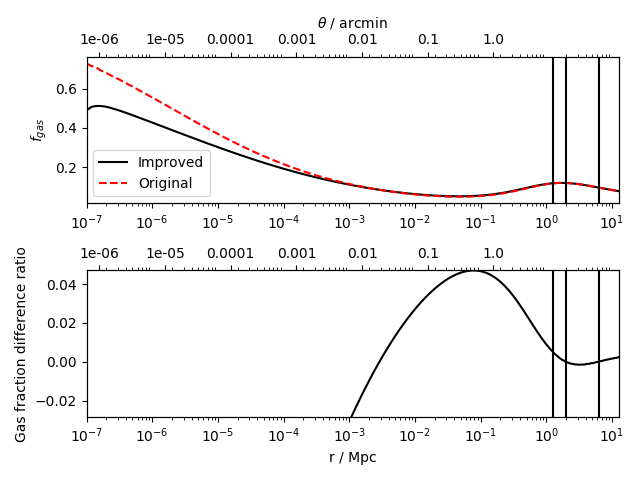}
  \caption{Comparison of the first-order and iteratively improved solutions for $M_{\rm g}(r)$ (left) and $f_{\rm gas}(r)$ (right), for a cluster at $z=0.15$ and with $M(r_{200}) = 1 \times10^{15} M_{\mathrm{Sun}}$.  Vertical lines show $r_{500}$, $r_{200}$ and the radius at which we cut off our cluster model.}
\label{Fi:mass_approx_demo}
  \end{center}
\end{figure*}

We can iteratively improve the solution for $\rho_{\rm g}$ by re-solving the hydrostatic equilibrium equation using updated estimates of $M(r) = M_{\rm g}(r) + M_{\rm dm}(r)$ at each iteration until the solutions converge. Figure~\ref{Fi:mass_approx_demo} shows the results of this improvement for a cluster at $z=0.15$ and with $M(r_{200}) = 1 \times10^{15} M_{\mathrm{Sun}}$ using an NFW profile for the dark matter; it is clear that the approximation works very well until small radii, and particularly near $r_{200}$ and $r_{500}$ which are where we use the model to solve for our normalisation factors, so we take the initial estimate of $M_{\rm g}(r)$ for our model. Note that a similar attempt at relaxing the mass approximation was attempted in \citet{KJThesis}, that involved solving differential equations in $M$ through the hydrostatic equilibrium relation. The work in \citet{KJThesis} gave the same results found in this paper.

Below we describe the specific implementation of PM~II, referring the reader to MO12 for more details of PM~I.


\subsubsection{Dark matter profiles}
\label{subsbusec:dmmodels}

Assuming an Einasto profile \citep{1965TrAlm...5...87E}, the dark matter density profile for a cluster $\rho_{\rm dm, PM \, II}$ is given by 
\begin{equation}
\label{eqn:einasto}
\rho_{\rm dm, PM \ II} = \rho_{-2} \exp \left[ -\frac{2}{\alpha_{\rm Ein}} \left( \left(\frac{r}{r_{-2}}\right)^{\alpha_{\rm Ein}} - 1 \right) \right],
\end{equation}
where $\alpha_{\rm Ein}$ is a shape parameter, $r_{-2}$ is the scale radius where the logarithmic derivative of the density is $-2$ (analogue to $r_{\rm s}$ in the NFW model, but note that in general $ r_{-2} \neq r_{\rm s}$), and $\rho_{-2}$ is the density at this radius. The parameter $\alpha_{\rm Ein}$ controls the degree of curvature of the profile. The larger its value, the more rapidly the slope varies with respect to $r$. In the limit that $\alpha_{\rm Ein} \rightarrow  0$, the logarithmic derivative is $-2$ for all $r$. For comparison we state the NFW dark matter profile used in PM I \citep{1995MNRAS.275..720N}
\begin{equation}
\label{eqn:nfw}
\rho_{\rm dm, PM \, I}(r) = \frac{\rho_{\rm s}}{\left(\frac{r}{r_{\rm s}}\right)\left(1+\frac{r}{r_{\rm s}}\right)^{2}},
\end{equation}
where $\rho_{\rm s}$ is a density normalisation constant, and $r_{\rm s}$ is another scale radius.
It is tempting to assume that the Einasto profile is capable of providing a better fit due to the fact that the Einasto profile has an extra degree of freedom (there are three for the Einasto profile, two for the NFW), the shape parameter. However \citet{2016MNRAS.457.4340K} claims that this is not strictly true, as the Einasto profile was seen to give a better fit to simulated dark matter haloes even with $\alpha_{\rm Ein}$ fixed. The asymptotic values of the logarithmic slope for the two profiles are as follows: as $r \rightarrow 0$ then $\mathrm{d}\ln \rho_{\mathrm{dm, PM \, I}}(r) / \mathrm{d}\ln r \rightarrow -1$ and $\mathrm{d}\ln \rho_{\mathrm{dm, PM \, II}}(r) / \mathrm{d}\ln r \rightarrow 0$. As $r \rightarrow \infty$ then $\mathrm{d}\ln \rho_{\mathrm{dm, PM \, I}}(r) / \mathrm{d}\ln r \rightarrow -3$ and $\mathrm{d}\ln \rho_{\mathrm{dm, PM \, II}}(r) / \mathrm{d}\ln r \rightarrow - \infty$. The magnitude of $\alpha_{\mathrm{Ein}}$ determines how quickly the slope changes between the two asymptotic values. Throughout this work when we refer to the NFW or Einasto model, we really mean the physical model which uses the NFW or Einasto model when considering the dark matter density profile. 
\\
Referring back to equation~\ref{eqn:einasto}, the ratio $r_{200}/r_{-2}$ is defined as the concentration parameter $c_{200}$ (and similarly $c_{200} = r_{200}/r_{\rm{s}}$ for the NFW profile). \citet{2014MNRAS.441.3359D} determine an analytical form for $c_{200}$ as a function of total mass and redshift (for the redshift range $z = [0,5]$) for Einasto profiles based on simulations similar to those described in \citet{2007MNRAS.378...55M} and \citet{2008MNRAS.391.1940M}
\begin{equation}
\label{eqn:einastoc200}
\log_{10}\left(c_{200}\right) = j(z) + k(z) \log_{10} \left[ \frac{M\left(r_{200}\right)}{10^{12}h^{-1}M_{\mathrm{Sun}}} \right],
\end{equation}
where $j(z) = 0.459 + 0.518\exp(-0.49z^{1.303})$ and $k(z) = -0.13 + 0.029z$. 

Following the method outlined above, we first calculate our first-order estimate for the total mass profile by integrating the Einasto profile
\begin{equation}
\label{eqn:massrho}
\begin{split}
M(r) &\approx \int_{0}^{r} 4\pi r'^{2} \rho_{\rm dm, PM \, II}(r') \, \rm{d}r' \\
     &= \frac{4 \pi \rho_{-2} r_{-2}^{3}}{\alpha_{\rm Ein}} \exp \left( 2 / \alpha_{\rm Ein} \right) \left( \frac{\alpha_{\rm Ein}}{2} \right) ^{3 / \alpha_{\rm Ein}} \\
     & \quad \times \gamma \left[ \frac{3}{\alpha_{\rm Ein}}, \frac{2}{\alpha_{\rm Ein}} \left( \frac{r}{r_{-2}} \right)^{\alpha_{\rm Ein}} \right],
\end{split}
\end{equation}
where $\gamma \left[a, x \right] = \int_{0}^{x} t^{a-1} e^{-t} \rm{d} t $ is the incomplete lower gamma function. The steps taken to get this result are given in Appendix~\ref{sec:einastointegral}. 
Equation~\ref{eqn:massrhocrit} can be evaluated at $r_{200}$ and equated with equation~\ref{eqn:massrho} evaluated at the same radius to obtain the following solution for $\rho_{-2}$
\begin{equation}
\label{eqn:rhom2}
\begin{split}
\rho_{-2} = & \frac{200}{3} \left(\frac{r_{200}}{r_{-2}} \right)^{3} \rho_{\rm crit}(z) \times \frac{1}{\left[1 / \alpha_{\rm Ein} \exp \left( 2 / \alpha_{\rm Ein} \right) \left( \frac{\alpha_{\rm Ein}}{2} \right) ^{3 / \alpha_{\rm Ein}}\right]} \\
          & \times \frac{1}{\gamma \left[ \frac{3}{\alpha_{\rm Ein}}, \frac{2}{\alpha_{\rm Ein}} \left( \frac{r_{200}}{r_{-2}} \right)^{\alpha_{\rm Ein}} \right]}.
\end{split}
\end{equation}
Equivalently, equation~\ref{eqn:massrho} can be evaluated at $r_{200}$ and set equal to the known value of $M(r_{200})$ to determine $\rho_{-2}$.  Note that $\rho_{-2}$ is the normalisation for our first-order approximation to the total mass profile; the corresponding normalisation for the dark matter mass profile is $(1 - f_{\rm gas}(r_{200})) \rho_{-2}$.

\begin{figure}
  \begin{center}
  \includegraphics[ width=0.90\linewidth]{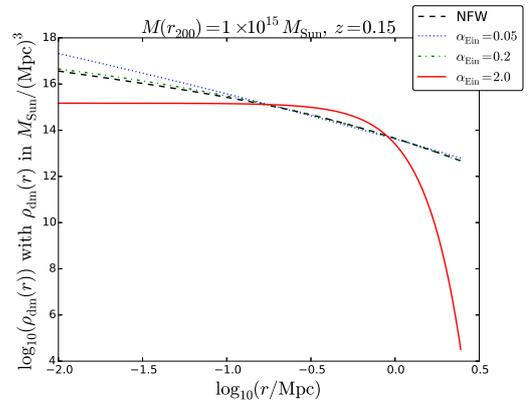}
  \caption{Logarithmic dark matter density profiles as a function of log cluster radius using NFW and Einasto models. Three values of the Einasto profile are used: $0.05, \, 0.2,$ and $2.0$. The additional input parameters used to generate these profiles are: $z =  0.15$, $M(r_{200}) = 1\times 10^{15} M_{\mathrm{Sun}}$ and $f_{\rm gas}(r_{200}) = 0.12$.}
\label{graph:einnfwdmdens}
  \end{center}
\end{figure}

Figure~\ref{graph:einnfwdmdens} shows the logarithmic dark matter density profiles as a function of $r$ for a cluster at $z =  0.15$ with $M(r_{200}) = 1\times 10^{15} M_{\mathrm{Sun}}$ and $f_{\rm gas}(r_{200}) = 0.12$ for PM I and PM II for the $\alpha_{\rm Ein}$ values: $0.05, \, 0.2, \, 2.0$. It is clear that the Einasto profiles diverge the most from each other at low $r$ and for the high $\alpha_{\rm Ein}$ value at high $r$ as well.


\subsubsection{Gas density and pressure profiles}
\label{subsubsec:pressuremodels}
As in PM I we follow \citet{2007ApJ...668....1N} and assume a generalised-NFW (GNFW) profile to parameterise the electron pressure $P_{\rm e}$ as a function of radius from the cluster centre,
\begin{equation}
\label{eqn:epressure}
P_{\rm e}(r) = \frac{P_{\rm ei}}{\left(\frac{r}{r_{\rm p}}\right)^{c}\left(1+\left(\frac{r}{r_{\rm p}}\right)^{a}\right)^{(b-c)/a}},
\end{equation}
where $P_{\rm ei}$ is the pressure normalisation constant and $r_{\rm p}$ is another characteristic radius, defined by $r_{\rm p} = r_{500}/c_{500}$. The parameters $a,\,b$ and $c$ describe the slope of the pressure profile at $ r \approx r_{\rm p}$, $r \gg r_{\rm p}$ and $r \ll r_{\rm p}$ respectively. The slope parameters are taken to be $a = 1.0510$, $b=5.4905$ and $c = 0.3081$. These `universal' values were taken from \citet{2010A&A...517A..92A} and are the best fit GNFW slope parameters derived from the REXCESS sub-sample (observed with XMM-Newton, \citealt{2007A&A...469..363B}), as described in Section~5 of Arnaud et al. We also take the Arnaud et al. value of the gas concentration parameter $c_{500}$ (note this is unrelated to the concentration parameter associated with the dark matter profile) which is $1.177$. Note that in MO12, KJ18 and KJ19 slightly different values derived for the standard self-similar case (Appendix~B of Arnaud et al.) were used ($a = 1.0620$, $b=5.4807$, $c = 0.3292$ and $c_{500}=1.156$). It was shown in \citet{2013MNRAS.430.1344O} that PM I is not affected by which of these two sets of parameters is used. \\
The analytical function used to convert from $r_{200}$ to $r_{500}$ in KJ19 is specific to the NFW dark matter profile case and so is not applicable in PM II. We have not found an analytic fitting function for the conversion in the case of an Einasto dark matter profile and so we obtain $r_{500}$ iteratively as described in Appendix~\ref{sec:r500newton}. 
 
We can relate the gas pressure $P_{\rm g}(r)$, to the electron pressure through the relation
\begin{equation}
\label{eqn:gaspressure}
\mu_{\rm g} P_{\rm g}(r) = \mu_{e} P_{\rm e}(r),
\end{equation}
where $\mu_{\rm e}$ is the mean gas mass per electron and $\mu_{\rm g}$ is the mean mass per gas particle. \citet{2000ApJ...540..614M} state that for a plasma with the cosmic helium mass fraction $C_{\rm{He}} = 0.24$ and the solar abundance values in \citet{1989GeCoA..53..197A}, then $\mu_{\rm e} = 1.146$ and $\mu_{\rm g} = 0.592$ in units of proton mass. \\
Incorporating equations~\ref{eqn:massrho} and~\ref{eqn:gaspressure} into the hydrostatic equilibrium equation gives the gas density
\begin{equation}
\label{eqn:rhog} 
\begin{split}
\rho_{\rm g}(r) =& \, \frac{\mu_{e}}{\mu_{\rm g}}\frac{P_{\rm ei}}{4\pi G \rho_{-2}r_{-2}^{3}} \frac{1}{\left[\left(1 / \alpha_{\rm Ein} \right) \exp( 2 / \alpha_{\rm Ein}) \left( \alpha_{\rm Ein} / 2 \right) ^{3 / \alpha_{\rm Ein}}\right]} \\
                 & \times \frac{r}{\gamma \left[ \frac{3}{\alpha_{\rm Ein}}, \frac{2}{\alpha_{\rm Ein}} \left( \frac{r_{200}}{r_{-2}} \right)^{\alpha_{\rm Ein}} \right]} \\
                 & \times \left(\frac{r}{r_{\rm p}}\right)^{-c}\bigg[1+\left(\frac{r}{r_{\rm p}}\right)^{a}\bigg]^{-\left(\frac{a+b-c}{a}\right)}\bigg[b\left(\frac{r}{r_{\rm p}}\right)^{a}+c\bigg].
\end{split}
\end{equation}
Note that even though an analytical expression for $\rho_{\rm g}$ exists, within this model there is no such equivalent for the total gas mass as
\begin{equation}
\label{eqn:gasmass}
M_{\rm g}(r) = \int_{0}^{r} 4\pi\rho_{\rm g}(r')r'^{2}\,\mathrm{d}r'
\end{equation}
must be integrated numerically. Hence $f_{\rm gas}(r) = M_{\rm g}(r)/M(r)$ does not have a closed form solution. Nevertheless, we can use equations~\ref{eqn:rhog} and~\ref{eqn:gasmass} to determine $P_{\rm ei}$ since we know $M(r_{200})$, $f_{\rm gas}(r_{200})$ and $r_{200}$. Evaluating equations~\ref{eqn:rhog} and~\ref{eqn:gasmass} at $r_{200}$ and solving for $P_{\rm ei}$ gives the following expression
\begin{equation}
\label{eqn:pei}
\begin{split}
 P_{\rm {ei}} = & \left(\frac{\mu_{\rm g}}{\mu_{e}}\right)
(G\rho_{-2}r^3_{-2})\left[\frac{\exp \left(2/\alpha_{\rm Ein} \right)}{\alpha_{\rm Ein}} \left(\alpha_{\rm Ein}/2\right)^{3/\alpha_{\rm Ein}} \right]M_{\rm g}(r_{200}) \\
              & \times  \frac{1}{ \bigint_{0}^{r_{200}} r'^{3}  \frac{\left[b \left(\frac{r'}{r_{\rm p}}\right)^{a} + c \right]}{\gamma\left[\frac{3}{\alpha_{\rm Ein}},\frac{2}{\alpha_{\rm Ein}} \left(\frac{r'}{r_{-2}}\right)^{\alpha_{\rm Ein}}\right]  \left(\frac{r'}{r_{\rm p}}\right)^c \left[1 + \left(\frac{r'}{r_{\rm p}}\right)^a\right]^{\left(\frac{a + b - c}{a}\right)} } {\rm d}r'},
\end{split}
\end{equation}
which must be evaluated numerically. Once $P_{\rm ei}$ has been calculated, the Comptonisation parameter as a function of projected radius on the sky can be calculated using equation~\ref{eqn:comptparamideal} which in turn can be used to calculate the surface brightness using equation~\ref{eqn:surfbrightcomplete}. Finally this can be Fourier transformed to get the quantity comparable to what an interferometer measures, so that the physical model can be used to analyse data obtained with AMI. 
\subsubsection{Additional cluster parameters}
\label{subsubsec:tempmodels}

The cluster gas properties are fully determined by the model, and so other parameters not used for AMI data analysis can readily be calculated.  For example, as stated in Section~2 of MO12 the radial profile of the electron number density is given by $n_{e}(r) = \rho_{\rm g}(r) / \mu_{e}$. Using the ideal gas assumption, the electron temperature $T_{e}(r)$ is given by
 
\begin{equation}
\label{eqn:tgas}
\begin{split}
T_{e}(r) = & \left(\frac{4\pi \mu_{\rm g} G\rho_{-2}r_{-2}^{3}}{k_{\rm B}}\right)
\left[\left(1/\alpha_{\rm Ein}\right) \exp \left(2/\alpha_{\rm Ein}\right) \left(\alpha_{\rm Ein}/2 \right)^{3/\alpha_{\rm Ein}}\right] \\
                   & \times \frac{\gamma\left[\frac{3}{\alpha_{\rm Ein}},\frac{2}{\alpha_{\rm Ein}}\left (\frac{r}{r_{-2}}\right)^ {\alpha_{\rm Ein}}\right]}{r} \\ 
                   & \times \left [1 + \left(\frac{r}{r_{\rm p}}\right)^{a} \right]\left[b \left(\frac{r}{r_{\rm p}}\right)^{a} + c \right]^{-1},
\end{split}
\end{equation}
which also equals $T_{\rm g}(r)$. This could be used to calculate relativistic corrections to the SZ signal.\\ 
The gas mass can be determined numerically from equation~\ref{eqn:gasmass},
\begin{equation}
\label{eqn:gasmass2}
\begin{split}
M_{\rm g}(r) =& \left(\frac{\mu_{e}}{\mu_{\rm g}}\right) \frac{1}{G} \frac{P_{\rm ei}}{\rho_{-2}} \frac{1}{\left[\left(1/\alpha_{\rm Ein}\right)\exp \left(2/\alpha_{\rm Ein} \right) \left(\alpha_{\rm Ein}/2\right)^{3/\alpha_{\rm Ein}}\right]r_{-2}^{3}} \\
              & \times \bigintss_{0}^{r} r'^{3} \frac{\left[b \left(\frac{r'}{r_{\rm p}}\right)^{a} + c \right]}{\gamma\left[\frac{3}{\alpha_{\rm Ein}},\frac{2}{\alpha_{\rm Ein}} \left(\frac{r'}{r_{-2}}\right)^{\alpha_{\rm Ein}}\right]} \\
              & \times \left(\frac{r'}{r_{\rm p}}\right)^c \left[1 + \left(\frac{r'}{r_{\rm p}}\right)^a\right]^{\left(\frac{a + b - c}{a}\right)} {\rm d}r',
\end{split}
\end{equation}
which can also be used to determine $f_{\rm gas}(r)$.


\subsection{Bayesian inference}

\label{subsec:bayesian}

The analysis of AMI data carried out in Section~\ref{subsec:results2} is done using Bayesian inference. We now give a summary of this in the context of both parameter estimation and model comparison.


\subsubsection{Parameter estimation}

Given a model $\mathcal{M}$ and data $\vec{\mathcal{D}}$, one can obtain probability distributions of the input parameters (also known as sampling parameters or model parameters) $\vec{\Theta}$ conditioned on $\mathcal{M}$ and $\vec{\mathcal{D}}$ using Bayes' theorem:
\begin{equation}\label{eqn:bayes}
Pr\left(\vec{\Theta}|\vec{\mathcal{D}},\mathcal{M}\right) = \frac{Pr\left(\vec{\mathcal{D}}|\vec{\Theta},\mathcal{M}\right)Pr\left(\vec{\Theta}|\mathcal{M}\right)}{Pr\left(\vec{\mathcal{D}}|\mathcal{M}\right)},
\end{equation}
where $Pr\left(\vec{\Theta}|\vec{\mathcal{D}},\mathcal{M}\right) \equiv \mathcal{P}\left(\vec{\Theta}\right)$ is the posterior distribution of the model parameter set, $Pr\left(\vec{\mathcal{D}}|\vec{\Theta},\mathcal{M}\right) \equiv \mathcal{L}\left(\vec{\Theta}\right)$ is the likelihood function for the data, $Pr\left(\vec{\Theta}|\mathcal{M}\right) \equiv \pi\left(\vec{\Theta}\right)$ is the prior probability distribution for the model parameter set, and $Pr\left(\vec{\mathcal{D}}|\mathcal{M}\right) \equiv \mathcal{Z}$ is the Bayesian evidence of the data given a model $\mathcal{M}$. The evidence can be interpreted as the factor required to normalise the posterior over the model parameter space:
\begin{equation}\label{eqn:evidence}
\mathcal{Z}\left(\vec{\mathcal{D}}\right) = \int \mathcal{L}\left(\vec{\Theta}\right) \pi\left(\vec{\Theta}\right)\, \mathrm{d}\vec{\Theta},
\end{equation} 
where the integral is carried out over the $N$-dimensional parameter space. For the models using AMI data considered here, the input parameter set can be split into two subsets, (which are assumed to be independent of one another): cluster parameters, $\vec{\Theta}_{\rm cl}$ and radio-source or `nuisance' parameters, $\vec{\Theta}_{\rm rs}$. The set of cluster parameters is $\alpha_{\rm Ein}$,  $M(r_{200})$, $f_{\rm gas}(r_{200})$, $z$, $x_{\rm c}$, and $y_{\rm c}$ (where the former only appears for PM II). $x_{\rm c}$ and $y_{\rm c}$ are the cluster centre offsets from the interferometer pointing centre, measured in arcseconds.  The cluster prior probability distributions are given in Section~\ref{subsubsec:priors}. For more details on the radio-source modelling, please refer to Section~5.2 of \citet{2009MNRAS.398.2049F} (from here on FF09). 
 For more information on the likelihood function and covariance matrix used in the AMI analysis, we refer the reader to \citet{2002MNRAS.334..569H} and Sections~5.3 of FF09 and~3.2.3 of KJ19.


\subsubsection{Model comparison}
\label{subsubsec:bayesmodel}
The nested sampling algorithm, \textsc{MultiNest} \citep{2009MNRAS.398.1601F} calculates $\mathcal{Z}\left(\vec{\mathcal{D}}\right)$ by making use of a transformation of the $N$-dimensional evidence integral into a one-dimensional integral. The algorithm also generates samples from $\mathcal{P}\left(\vec{\Theta}\right)$ as a by-product, meaning that it is suitable for both the parameter estimation and model comparison aspects of this work. Comparing models in a Bayesian way can be done as follows. The probability of a model $\mathcal{M}$, conditioned on $\vec{\mathcal{D}}$ can also be calculated using Bayes' theorem
\begin{equation}\label{eqn:bayesmodel}
Pr\left(\mathcal{M}|\vec{\mathcal{D}}\right) = \frac{Pr\left(\vec{\mathcal{D}}|\mathcal{M}\right)Pr\left(\mathcal{M}\right)}{Pr\left(\vec{\mathcal{D}}\right)}.
\end{equation}
Hence for two models $\mathcal{M}_{1}$ and $\mathcal{M}_{2}$, the ratio of the probability of the models conditioned on the same dataset is given by 
\begin{equation}\label{eqn:bayesmodelcomparison}
\frac{Pr\left(\mathcal{M}_{1}|\vec{\mathcal{D}}\right)}{Pr\left(\mathcal{M}_{2}|\vec{\mathcal{D}}\right)} = \frac{Pr\left(\vec{\mathcal{D}}|\mathcal{M}_{1}\right)Pr\left(\mathcal{M}_{1}\right)}{Pr\left(\vec{\mathcal{D}}|\mathcal{M}_{2}\right)Pr\left(\mathcal{M}_{2}\right)},
\end{equation}
where $Pr(\mathcal{M}_{2}) / Pr(\mathcal{M}_{1})$ is the a-priori probability ratio of the models. We set this to one, i.e. place no bias towards a particular model before performing the analysis. Hence the ratio of the probabilities of the models given the data is equal to the ratio of the evidence values obtained from the respective models (for brevity we define $\mathcal{Z}_{i} \equiv Pr\left(\vec{\mathcal{D}}|\mathcal{M}_{i}\right)$). 
The evidence is simply the average of the likelihood function over the parameter space, weighted by the prior distribution. This means that the evidence is larger for a model if more of its parameter space is likely and smaller for a model with large areas in its parameter space having low likelihood values. A larger parameter space, either in the form of higher dimensionality or a wider domain results in a lower evidence value all other things being equal. Hence the evidence `punishes' more complex models over basic (lower dimensionality / smaller input parameter space domains) ones which give an equally good fit to the data. Thus the evidence automatically implements Occam's razor: when you have two competing theories that make exactly the same predictions, the simpler one is the preferred. \citet{jeffreys} provides a scale for interpreting the ratio of evidences as a means of performing model comparison (see Table~\ref{tab:jeffreys}). A value of $\ln (\mathcal{Z}_{1} / \mathcal{Z}_{2})$ above $5.0$ (less than $-5.0$) presents strong evidence in favour of $\mathcal{M}_{1}$ ($\mathcal{M}_{2}$). Values $ 2.5 \leq \ln (\mathcal{Z}_{1} / \mathcal{Z}_{2}) < 5.0$ ($ -5.0 < \ln (\mathcal{Z}_{1} / \mathcal{Z}_{2}) \leq -2.5 $) present moderate evidence in favour of $\mathcal{M}_{1}$ ($\mathcal{M}_{2}$). Values $ 1 \leq \ln (\mathcal{Z}_{1} / \mathcal{Z}_{2}) < 2.5$ ($ -2.5 < \ln (\mathcal{Z}_{1} / \mathcal{Z}_{2}) \leq -1 $) present weak evidence in favour of $\mathcal{M}_{1}$ ($\mathcal{M}_{2}$). Finally, values $ -1.0 < \ln (\mathcal{Z}_{1} / \mathcal{Z}_{2}) < 1.0 $ require more information to come to a conclusion over preference between $\mathcal{M}_{1}$ and $\mathcal{M}_{2}$.

\begin{table*}
\begin{center}
\begin{tabular}{{l}{c}{c}}
\hline
$\ln (\mathcal{Z}_{1} / \mathcal{Z}_{2})$  & Interpretation & Probability of favoured model \\ 
\hline
$\leq 1.0$ & better data are needed & $\leq 0.75$ \\
$\leq 2.5 $ & weak evidence in favour of $\mathcal{M}_{1}$ & $0.923$ \\
$\leq 5.0$ & moderate evidence in favour of $\mathcal{M}_{1}$ & $0.993$ \\
$ > 5.0$ & strong evidence in favour of $\mathcal{M}_{1}$ & $ > 0.993 $ \\
\hline
\end{tabular}
\caption{Jeffreys scale for assessing model preferability based on the log of the evidence ratio of two models $\mathcal{M}_{1}$ and $\mathcal{M}_{2}$.}\label{tab:jeffreys}
\end{center}
\end{table*}


\subsubsection{Prior probability distributions}
\label{subsubsec:priors}
For both PM I and PM II we adopt the following approach (excluding any mention of $\alpha_{\rm Ein}$ in the former case). \\
Following FF09, the cluster parameters are assumed to be independent of one another, so that
\begin{equation}\label{eqn:cluspriors}
\pi(\vec{\Theta}_{\rm cl}) = \pi(\alpha_{\rm Ein})\pi(M(r_{200}))\pi(f_{\rm gas}(r_{200}))\pi(z)\pi(x_{\rm c})\pi(y_{\rm c}).
\end{equation} 
Table~\ref{tab:clusterpriors} lists the type of prior used for each cluster parameter and the probability distribution parameters. The values used for $z$ and $\alpha_{\rm Ein}$ will be specified on a case by case basis in Section~\ref{subsec:results2}.  The $f_{\rm gas}(r_{200})$ prior is based on \citet{2011ApJS..192...18K}.  We note that more recent observations support a higher value, e.g.\ \citet{2019A&A...621A..40E} find a median $f_{\rm gas}(r_{200}) = 0.146$ for a sample of high-mass, low-redshift clusters; a more correct prior should take account mass-dependence (e.g. \citealt{2015MNRAS.450..896D}) but we leave this refinement for future work.

\begin{table}
\begin{center}
\begin{tabular}{{l}{c}}
\hline
Parameter & Prior distribution \\ 
\hline
$x_{\rm c}$ & $\mathcal{N}(0'', 60'')$ \\
$y_{\rm c}$ & $\mathcal{N}(0'', 60'')$ \\
$z$ & $\delta(z)$ \\
$M(r_{200})$ & $\mathcal{U} [ \log (0.5\times 10^{14} M_{\mathrm{Sun}}), \log (50\times 10^{14} M_{\mathrm{Sun}})]$ \\
$f_{\rm gas}(r_{200})$ & $\mathcal{N}(0.12, 0.02)$ \\
$\alpha_{\rm Ein}$ & $\delta(\alpha_{\rm Ein})$ \\
\hline
\end{tabular}
\caption{Cluster parameter prior distributions, where the normal distributions are parameterised by their mean and standard deviations.}
\label{tab:clusterpriors}
\end{center}
\end{table}


\section{Results}
\label{sec:results}


\subsection{Cluster parameter profiles}
\label{subsec:results1}

We first present the results of using the Einasto model in the profiling of cluster dark matter for a range of different cluster input parameters, along with the equivalent results from PM I. \\
We consider two input masses, $M(r_{200}) = 1\times 10^{14} M_{\mathrm{Sun}}$ and $M(r_{200}) = 1\times 10^{15} M_{\mathrm{Sun}}$ 
and use $z$-values of $0.15$ and $0.9$.  We take $f_{\rm gas}(r_{200}) = 0.12$ following \citet{2011ApJS..192...18K}, and consider $\alpha_{\rm Ein}$ values of $0.05, \, 0.2,$ and $ 2.0$ -- see Figure~\ref{graph:einnfwdmdens}. \citet{2016MNRAS.457.4340K} find a positive correlation between $\alpha_{\rm Ein}$ and the mass of a cluster, suggesting that the clusters considered here with relatively large (small) values for $\alpha_{\rm Ein}$ and small (large) values for $M(r_{200})$ may be considered unphysical based on the findings of their paper. Nevertheless we proceed with our range of clusters as we would like to analyse the behaviour at these extreme values. We note that the same $r$ range ($-2 \leq \log_{10}(r) \leq 0.5$ (where $r$ is in units of Mpc)) is considered for each cluster, and thus even though each parameter profile is self-similar in $r$ with respect to mass and redshift, they are different for each cluster over the range of $r$ considered here. 


\subsubsection{Dark matter mass profiles}
\label{subsubsec:dmmresults}

\begin{figure*}
  \begin{center}
    \begin{tabular}{@{}cc@{}}
     \includegraphics[ width=0.5\linewidth]{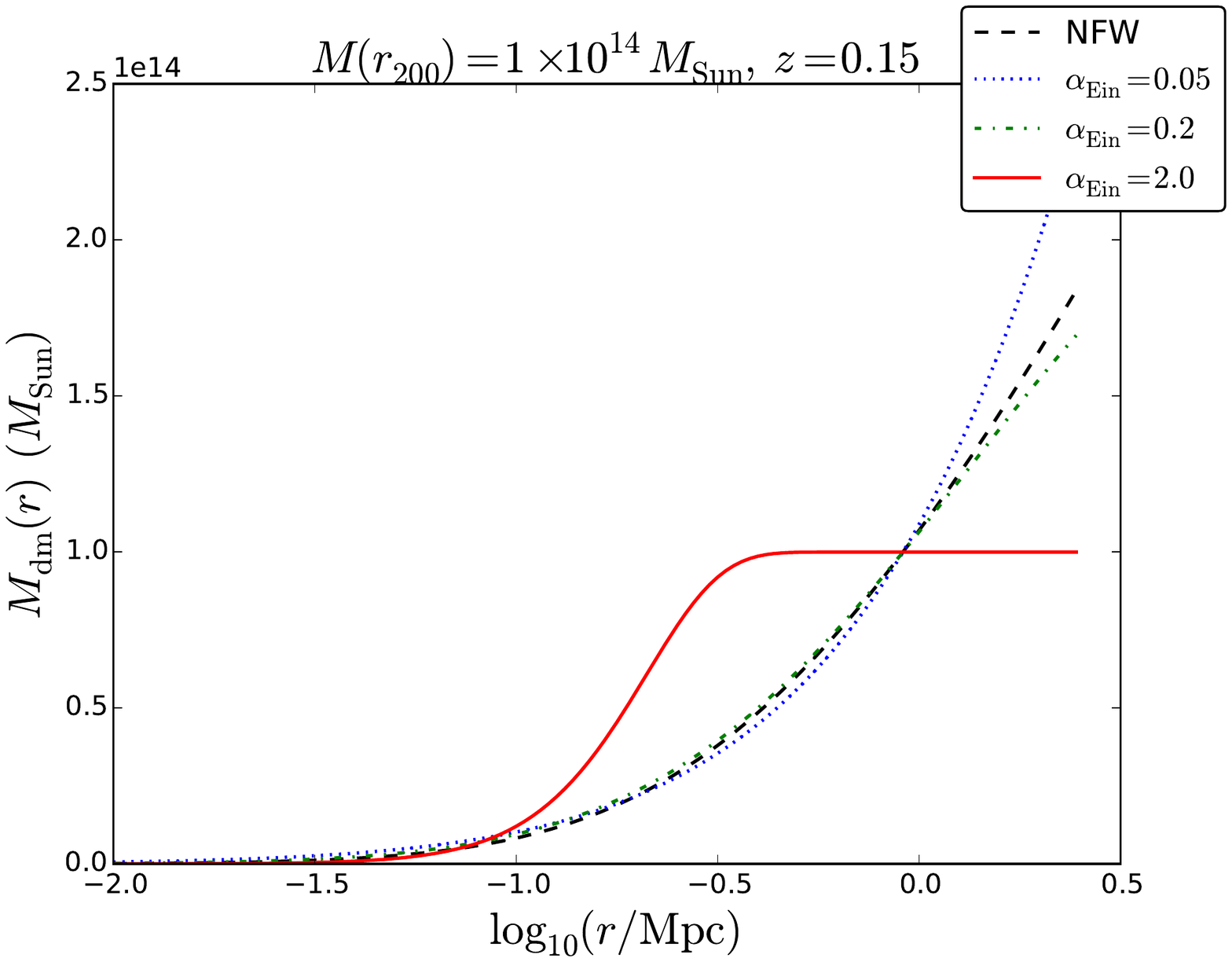} &
     \includegraphics[ width=0.5\linewidth]{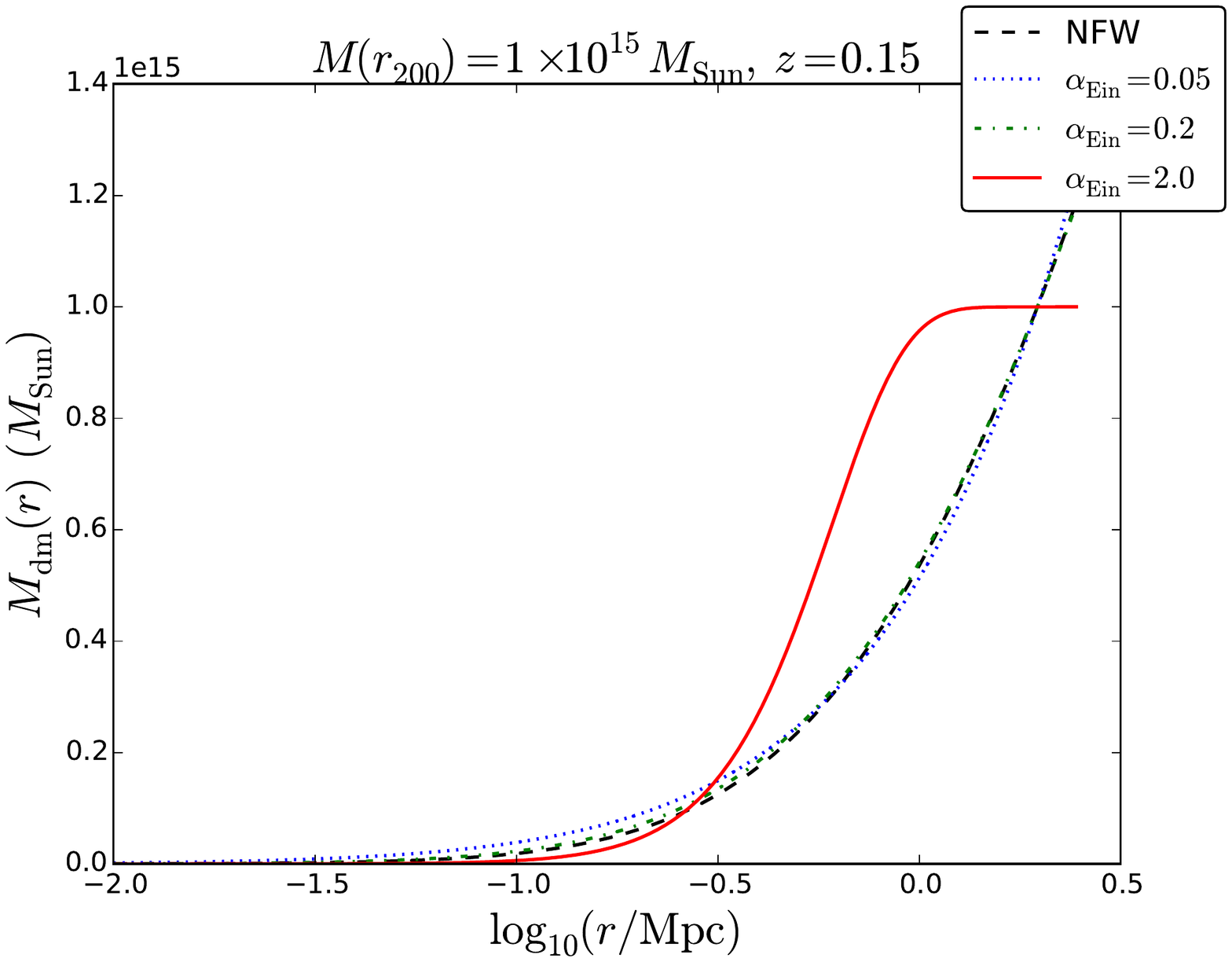} \\
     \includegraphics[ width=0.5\linewidth]{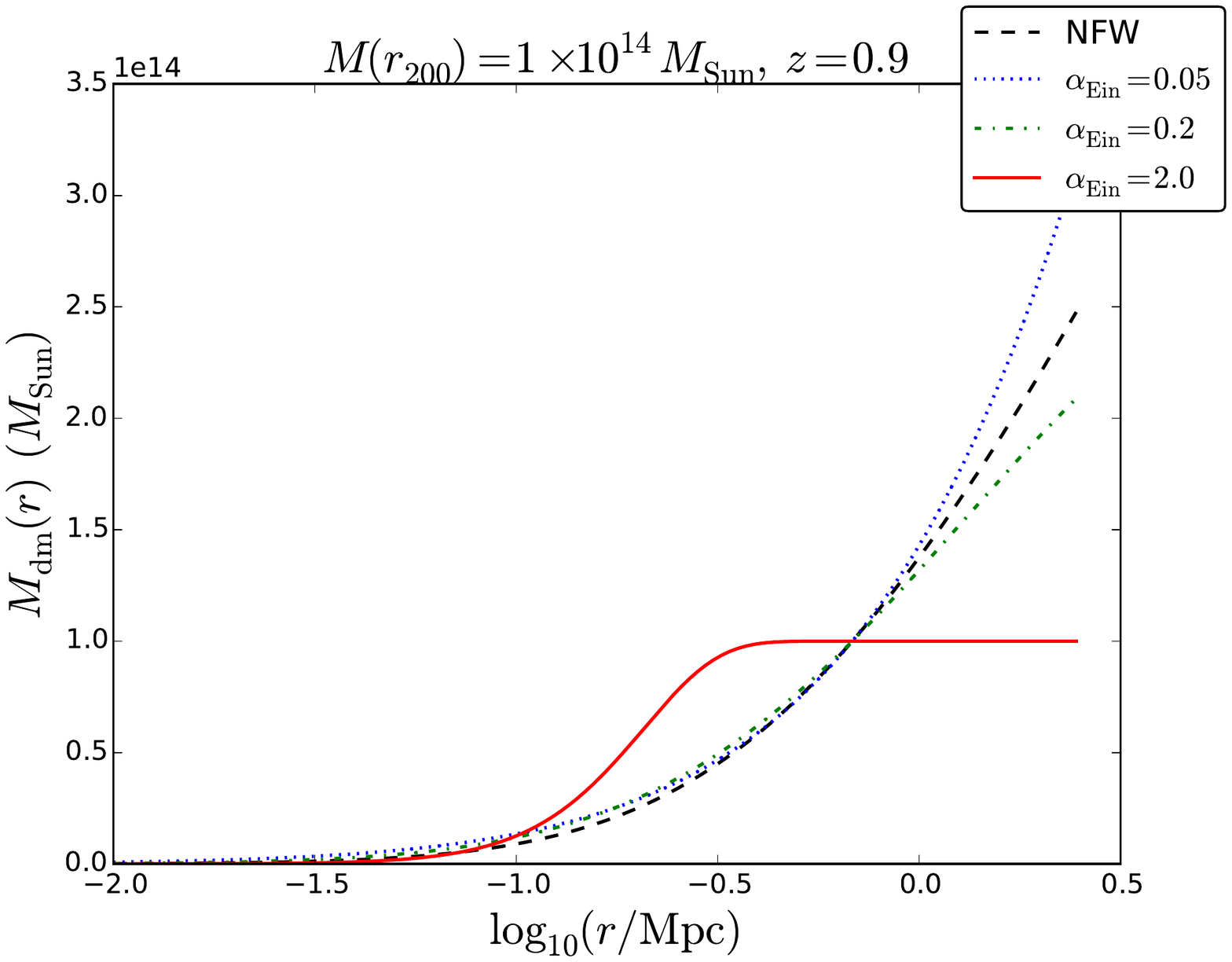} &
     \includegraphics[ width=0.5\linewidth]{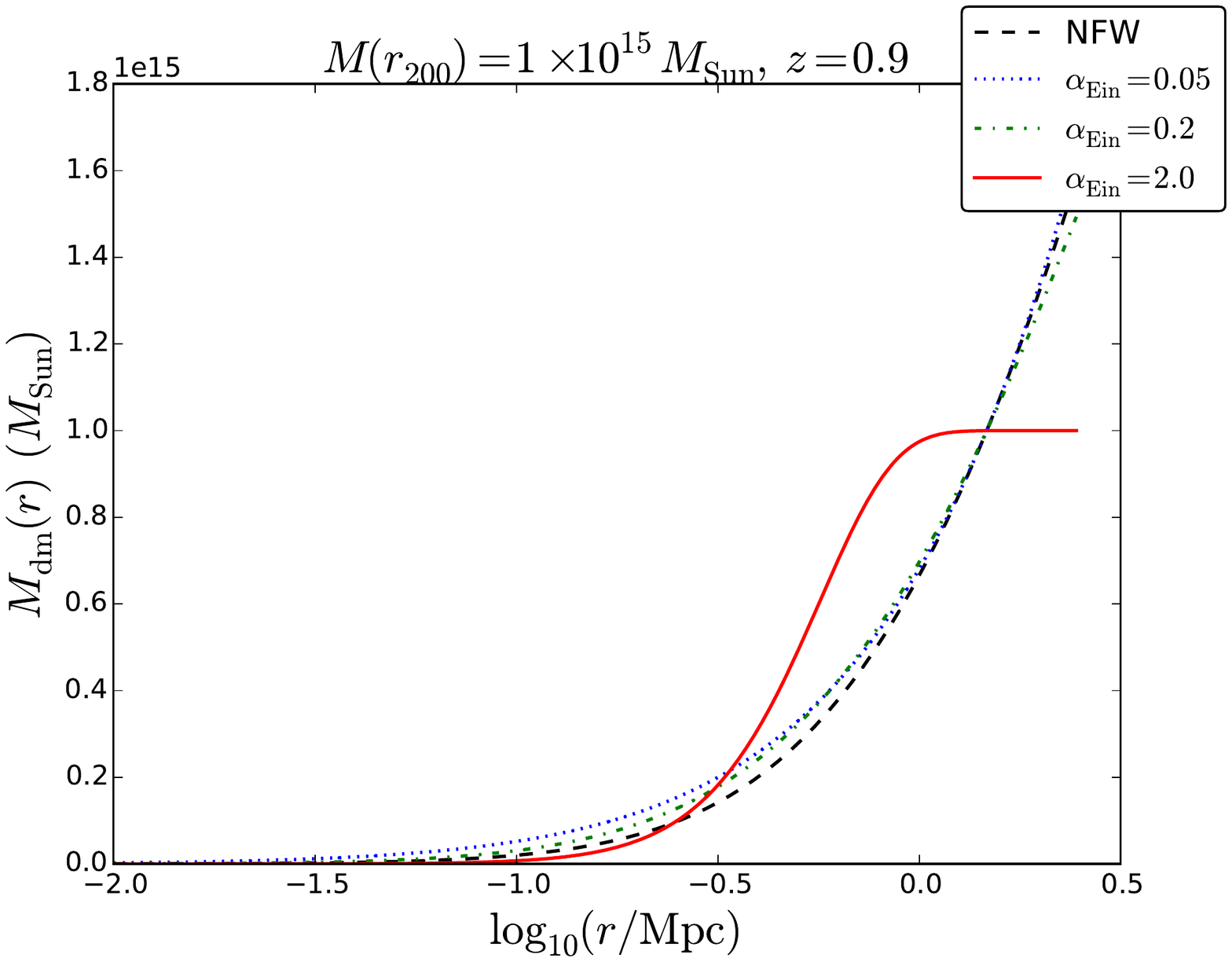} \\
    \end{tabular}

  \caption{Dark matter mass profiles as a function of log cluster radius using NFW and Einasto models. Values of $\alpha_{\mathrm{Ein}} = 0.05$, $0.2$, and $2.0$ are used as inputs. Top row has $z = 0.15$, bottom row has $z = 0.9$. Left column has $M(r_{200}) = 1\times 10^{14} M_{\mathrm{Sun}}$, right column has $M(r_{200}) = 1\times 10^{15} M_{\mathrm{Sun}}$.}
\label{graph:einnfwdmmass}
  \end{center}
\end{figure*}

Figure~\ref{graph:einnfwdmmass} shows the dark matter mass profiles. The Einasto profiles are calculated using equation~\ref{eqn:massrho} and the NFW profile from the equivalent relation given in MO12 (equation~5). 
Note that these are proportional to the first-order total mass solutions, i.e.\ $M(r) \approx (1 - f_{\rm gas}(r_{200})) M_{\rm dm}(r)$.

The $\alpha_{\rm Ein} = 2$ case always converges quickly as the density rapidly falls to zero, while the other three profiles including the NFW show divergent behaviour at the largest radii considered here. The high mass inputs result in similar profiles for the $\alpha_{\rm Ein} = 0.05$, $\alpha_{\rm Ein} = 0.2$ and NFW cases, whereas the low mass inputs result in the $\alpha_{\rm Ein} = 0.05$ case diverging somewhat more rapidly than the others.

\subsubsection{Gas density profiles}
\label{subsubsec:grhoresults}

\begin{figure*}
  \begin{center}
    \begin{tabular}{@{}cc@{}}
     \includegraphics[ width=0.5\linewidth]{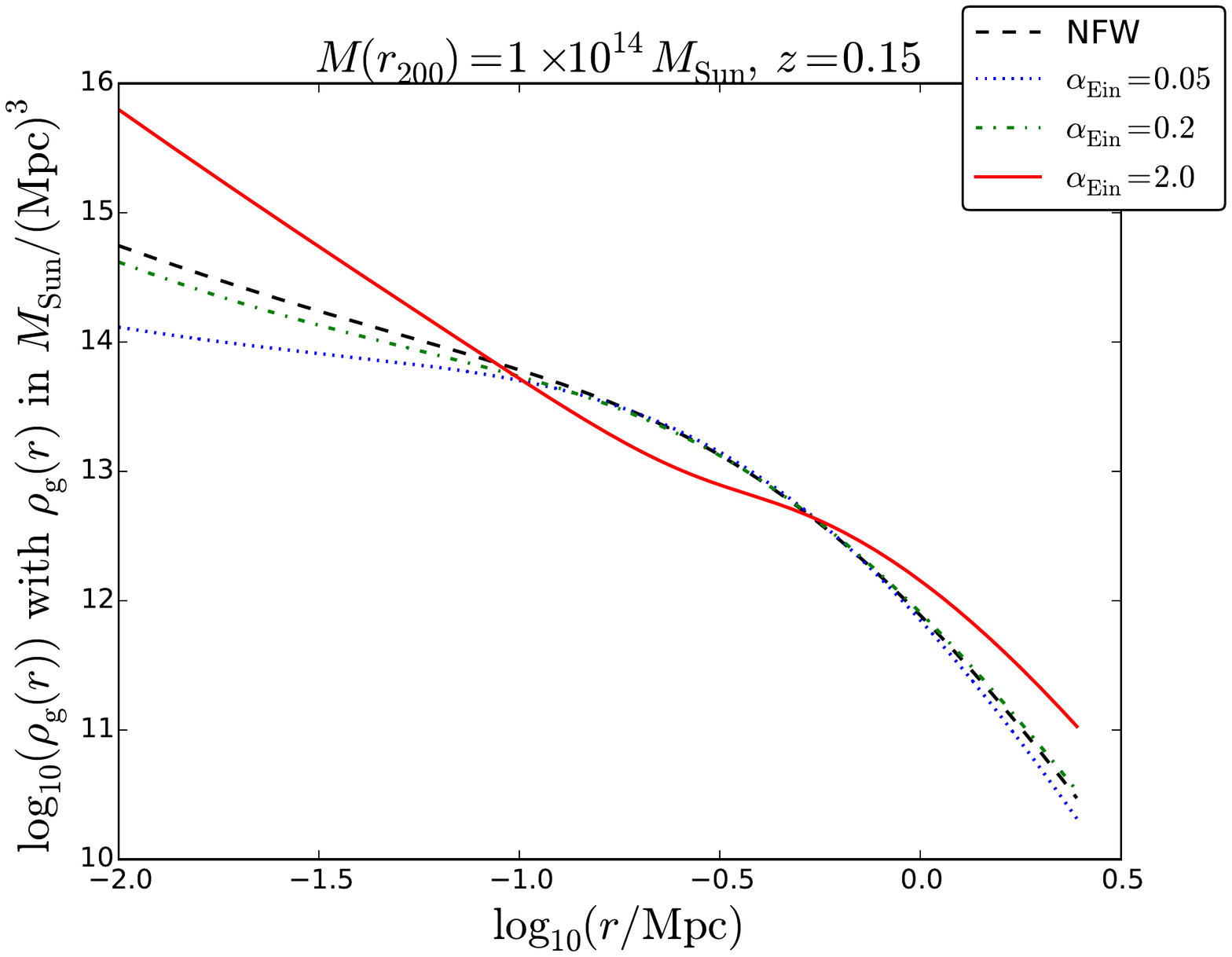} &
     \includegraphics[ width=0.5\linewidth]{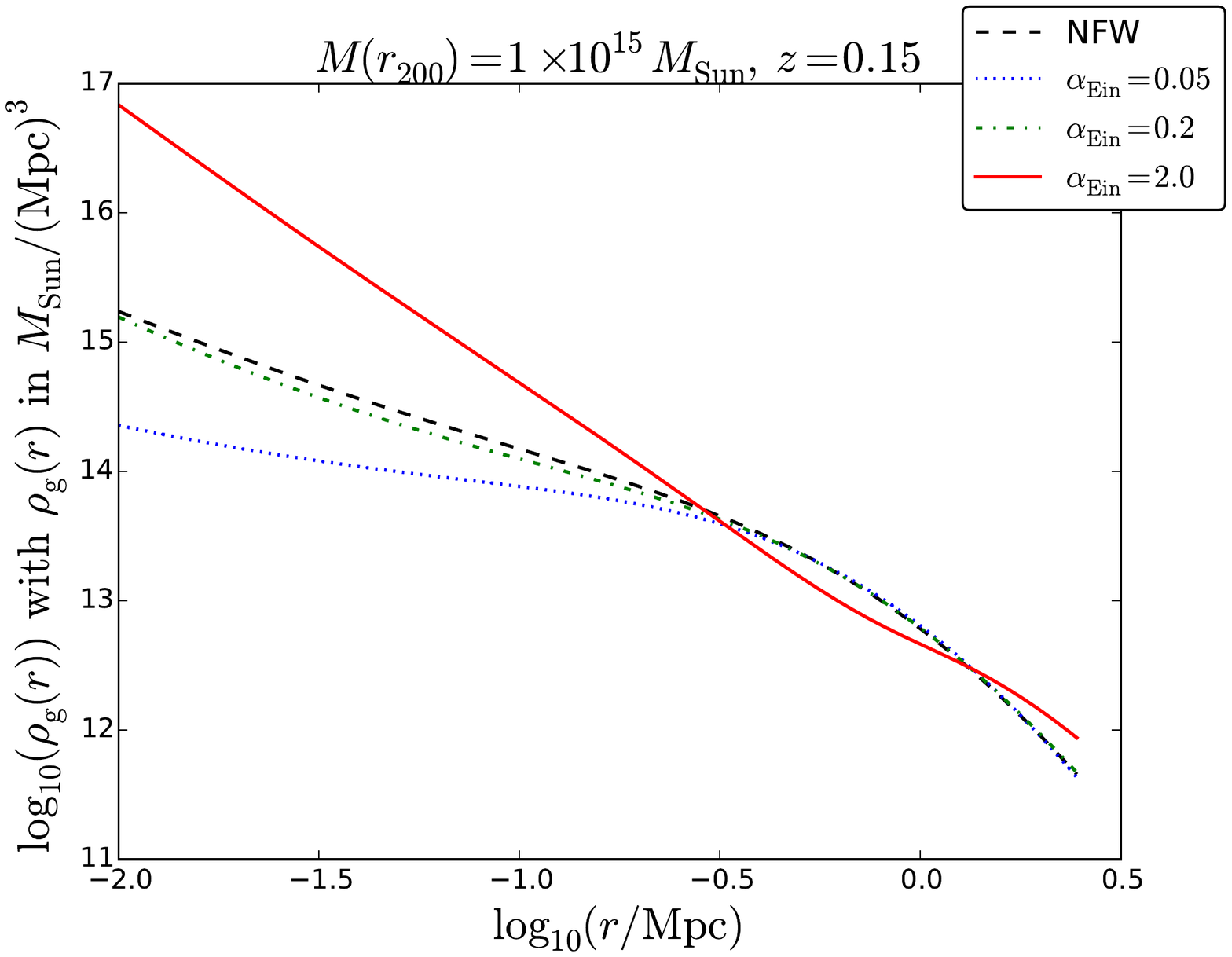} \\
     \includegraphics[ width=0.5\linewidth]{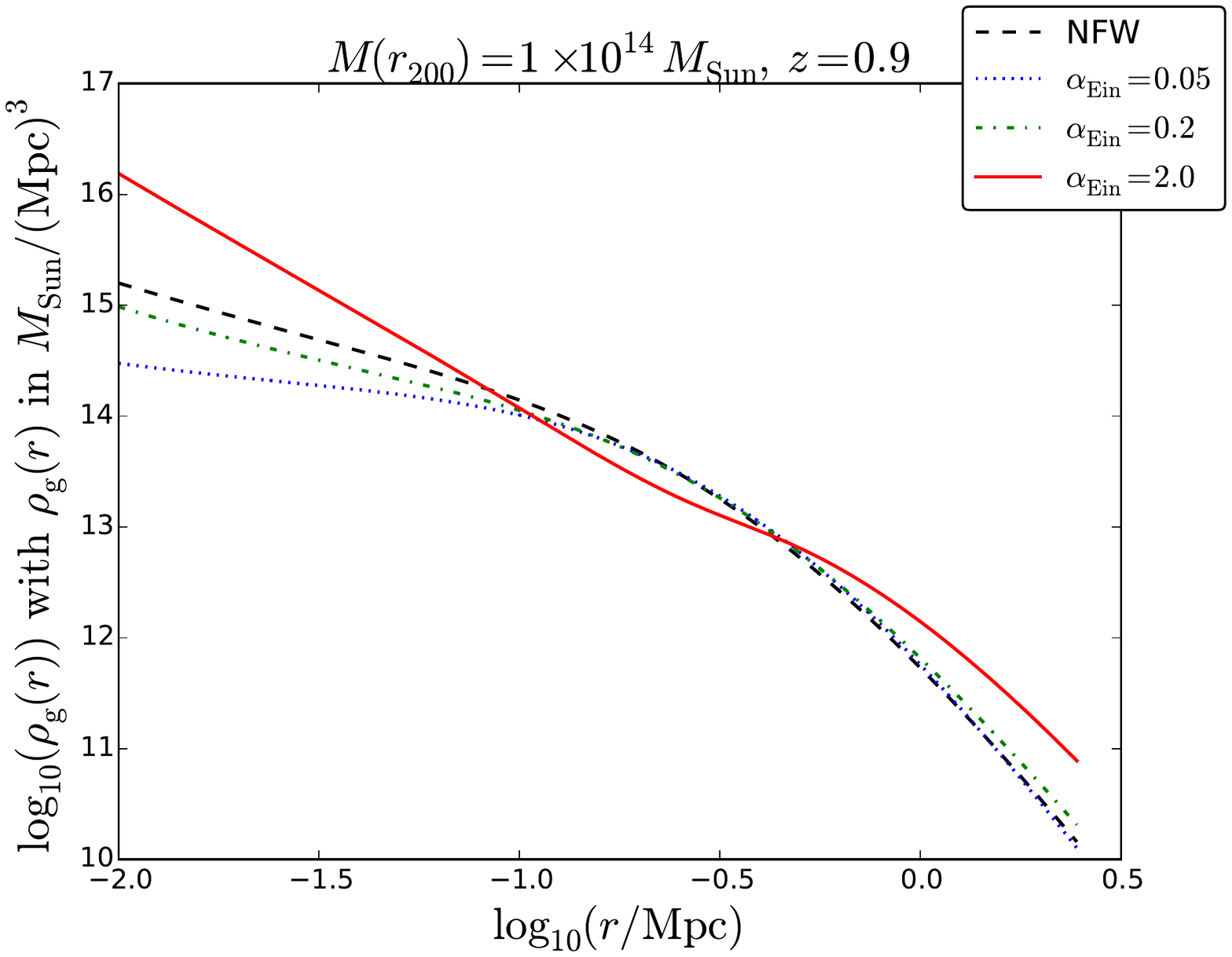} &
     \includegraphics[ width=0.5\linewidth]{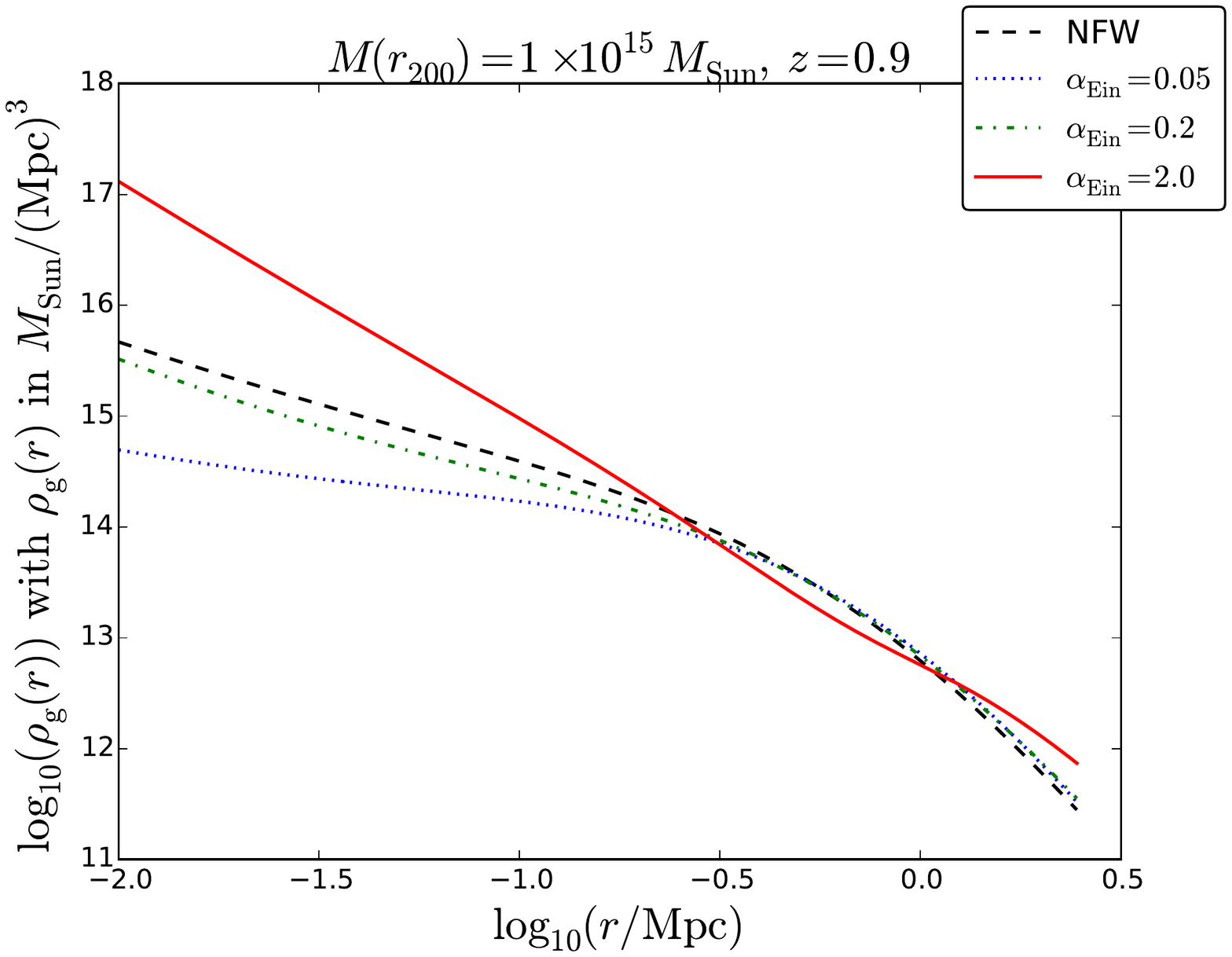} \\
    \end{tabular}
  \caption{Logarithmic gas density profiles as a function of log cluster radius using NFW and Einasto models. Values of $\alpha_{\mathrm{Ein}} = 0.05$, $0.2$, and $2.0$ are used as inputs. Top row has $z = 0.15$, bottom row has $z = 0.9$. Left column has $M(r_{200}) = 1\times 10^{14} M_{\mathrm{Sun}}$, right column has $M(r_{200}) = 1\times 10^{15} M_{\mathrm{Sun}}$.}
\label{graph:einnfwgrho}
  \end{center}
\end{figure*}

Figure~\ref{graph:einnfwgrho} shows the gas density profiles. The Einasto profiles are calculated using equation~\ref{eqn:rhog} and the NFW profile from the equivalent relation given in MO12 (equation~6). 
Note that these are the first-order solutions for $\rho_{\rm g}(r)$, i.e.\ assuming $M(r) \propto M_{\rm dm}(r)$ when solving the hydrostatic equilibrium equation.

The plots show that the profiles are similar for all inputs of mass and redshift, with the $\alpha_{\rm Ein} = 0.2$ Einasto profile again most resembling the NFW profile. However, the $\alpha_{\rm Ein} = 2.0$ profile has the highest gas density at high $r$ for both masses and both $z$ values.

\subsubsection{Gas mass profiles}
\label{subsubsec:gmassresults}
\begin{figure*}
  \begin{center}
  \begin{tabular}{@{}cc@{}}
     \includegraphics[ width=0.5\linewidth]{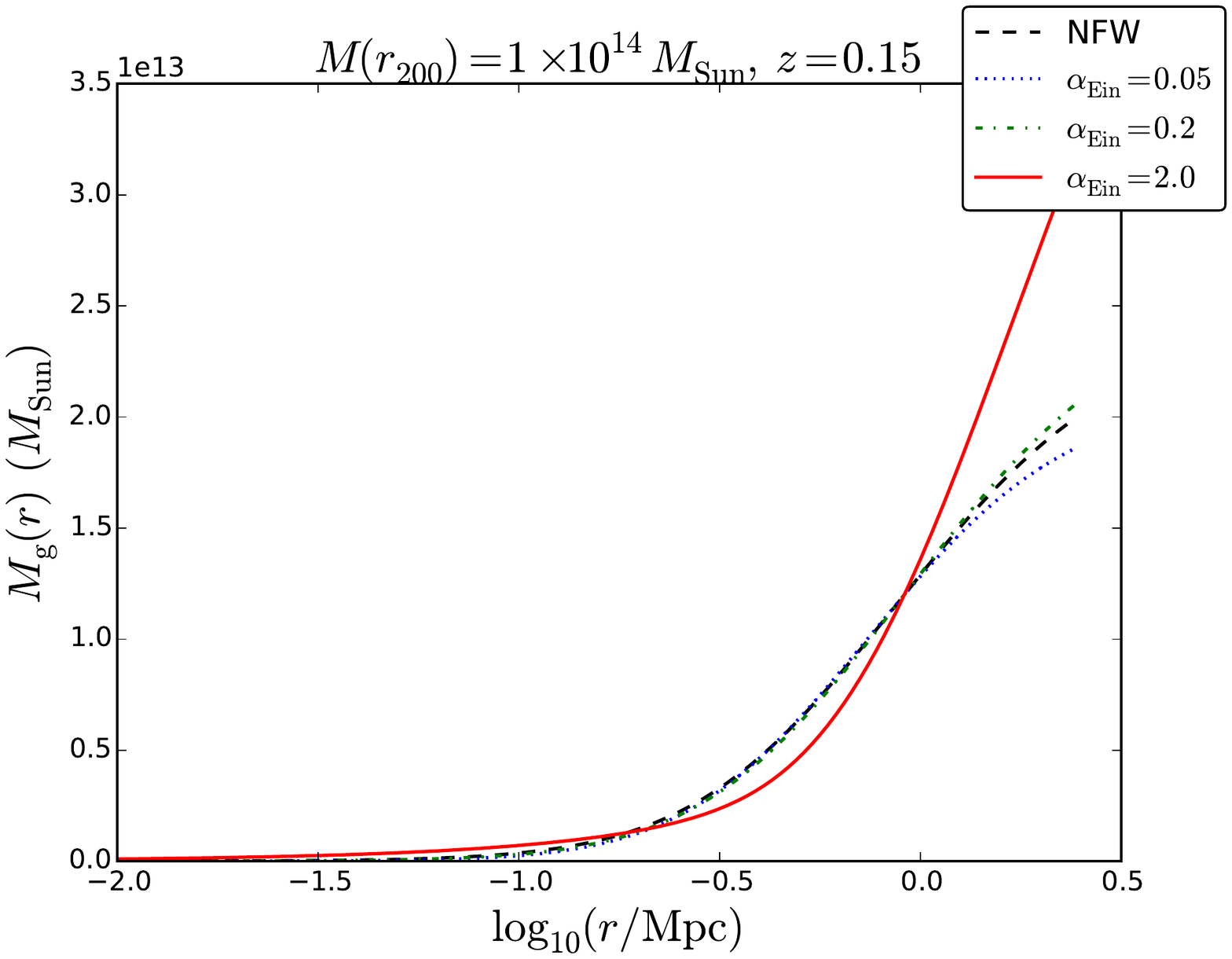} &
     \includegraphics[ width=0.5\linewidth]{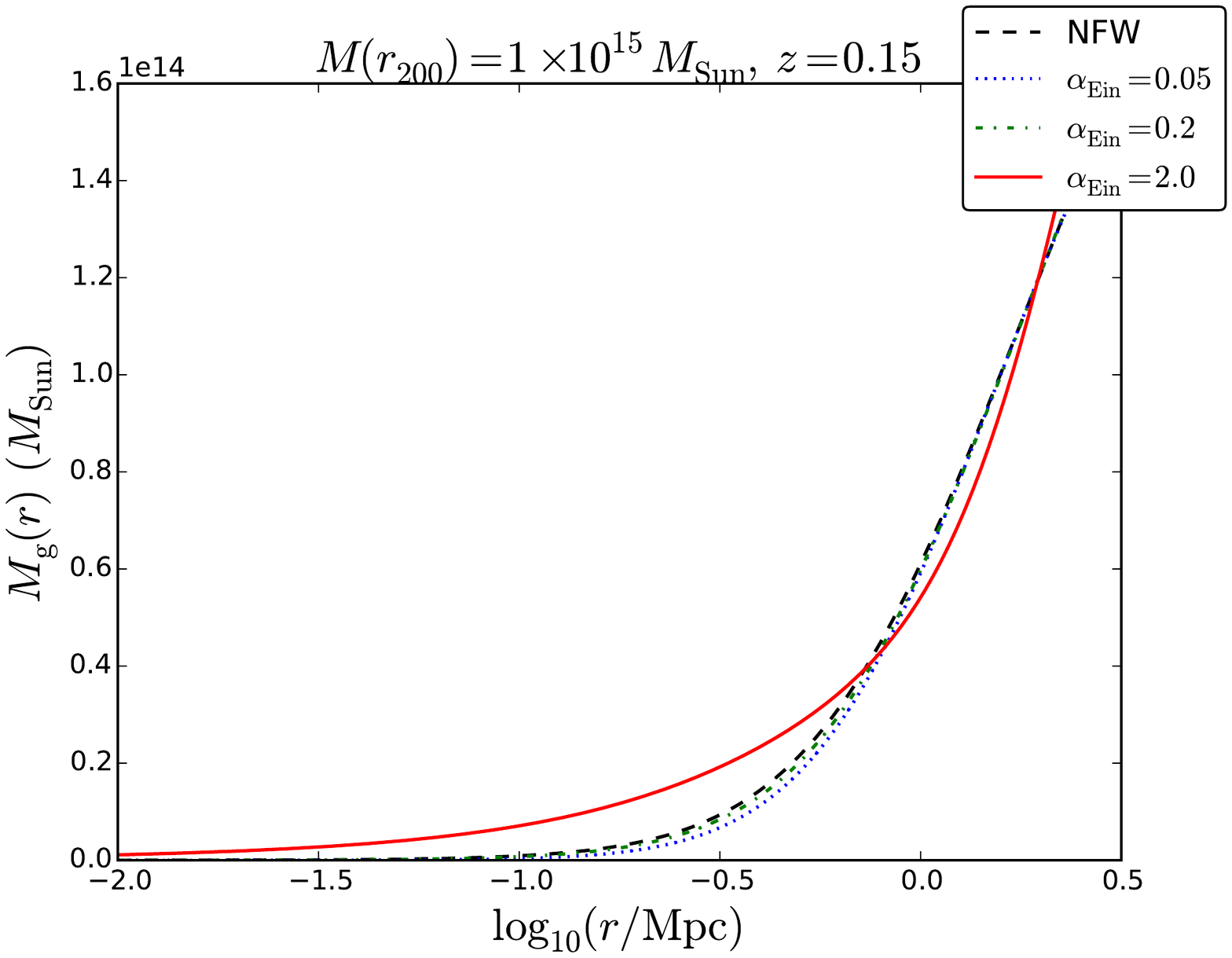} \\
     \includegraphics[ width=0.5\linewidth]{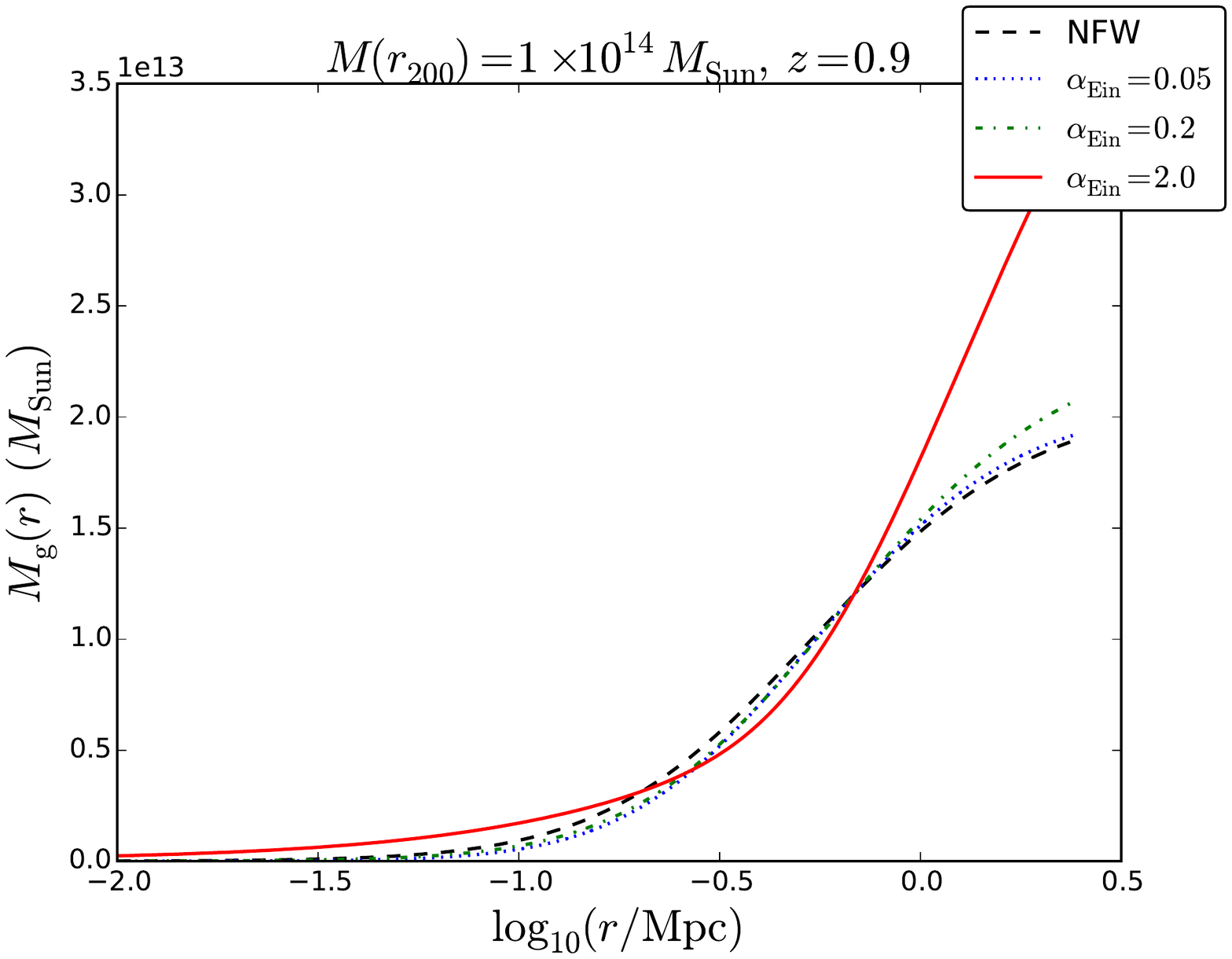} &
     \includegraphics[ width=0.5\linewidth]{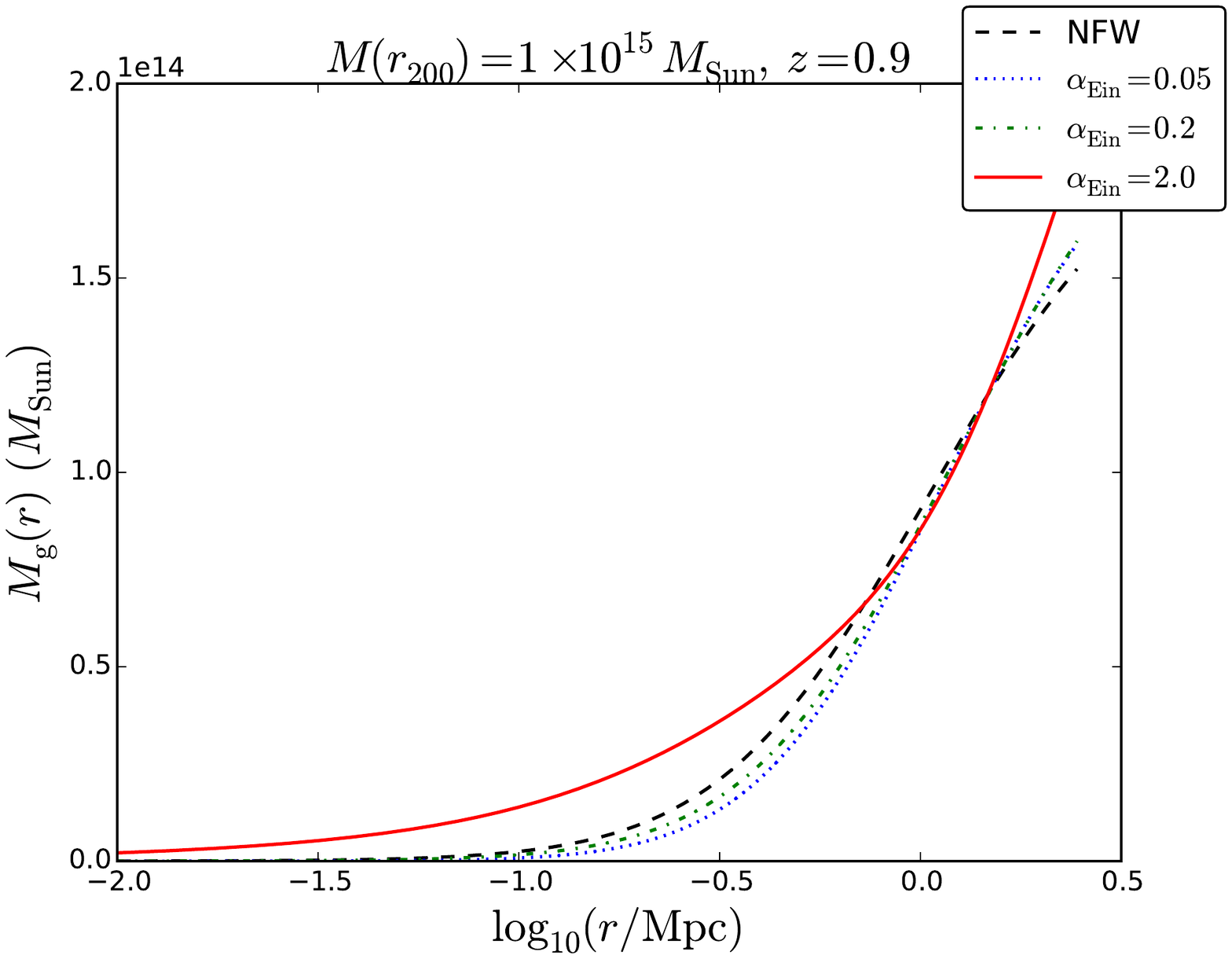} \\
    \end{tabular}
  \caption{Gas mass profiles as a function of log cluster radius using NFW and Einasto models. Values of $\alpha_{\mathrm{Ein}} = 0.05$, $0.2$, and $2.0$ are used as inputs. Top row has $z = 0.15$, bottom row has $z = 0.9$. Left column has $M(r_{200}) = 1\times 10^{14} M_{\mathrm{Sun}}$, right column has $M(r_{200}) = 1\times 10^{15} M_{\mathrm{Sun}}$.}
\label{graph:einnfwgm}
  \end{center}
\end{figure*}

Figure~\ref{graph:einnfwgm} shows $M_{\rm g}(r)$ as a function of cluster radius. As in Figure~\ref{graph:einnfwdmmass} with the dark matter mass profiles, the high mass inputs correspond to divergent behaviour at large $r$. But for $\alpha_{\rm Ein} = 2.0$ the profile of $M_{\rm g}(r)$ also shows a more noticeable such divergence.
Furthermore, in all four input parameter cases, $\alpha_{\rm Ein} = 2.0$ shows more divergent behaviour than other values of $\alpha_{\rm Ein}$ and the NFW profile in gas mass, which is in contrast to the dark matter mass profiles. 

\subsubsection{Gas temperature profiles}
\label{subsubsec:gtempresults}
\begin{figure*}
  \begin{center}
  \begin{tabular}{@{}cc@{}}
     \includegraphics[ width=0.5\linewidth]{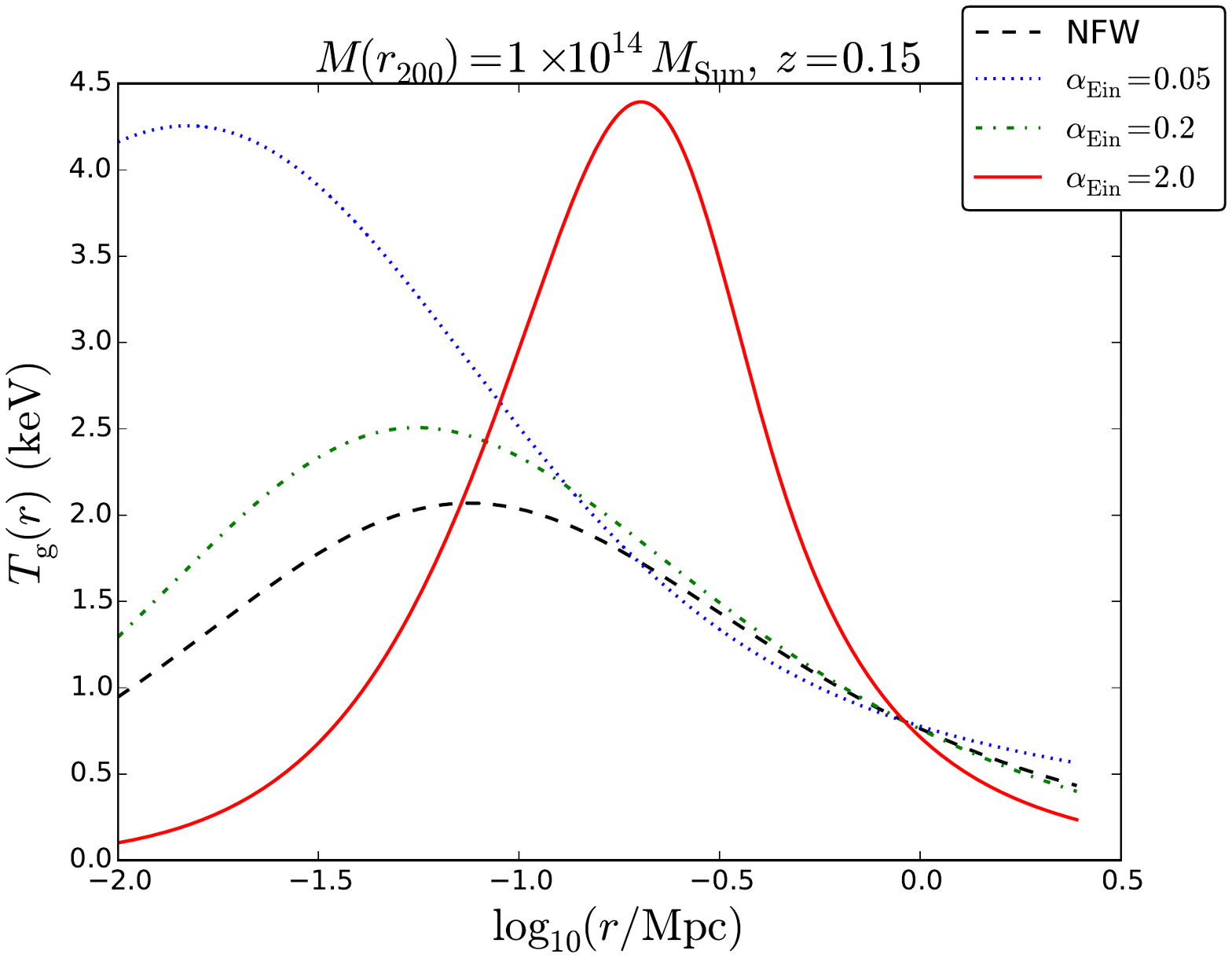} &
     \includegraphics[ width=0.5\linewidth]{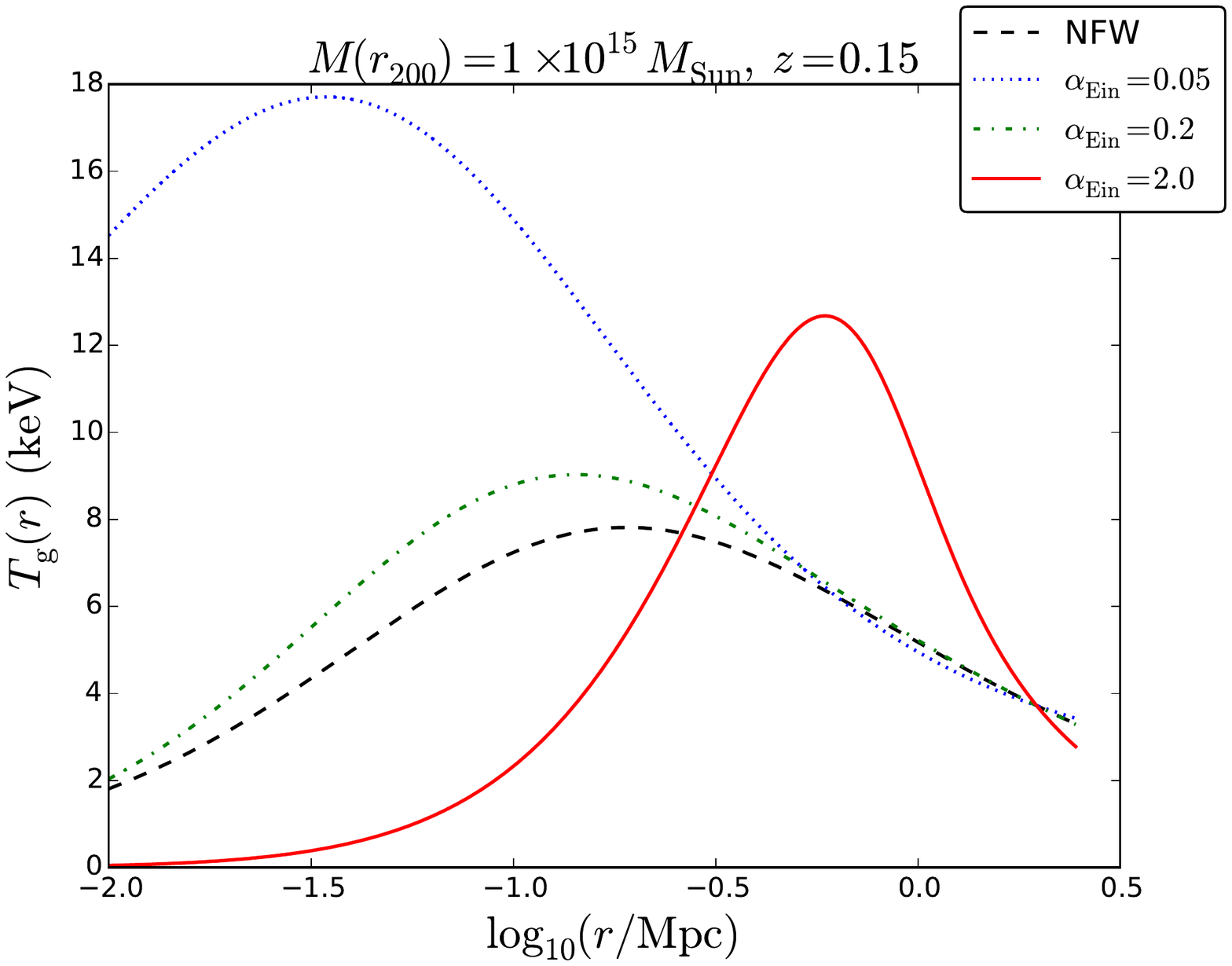} \\
     \includegraphics[ width=0.5\linewidth]{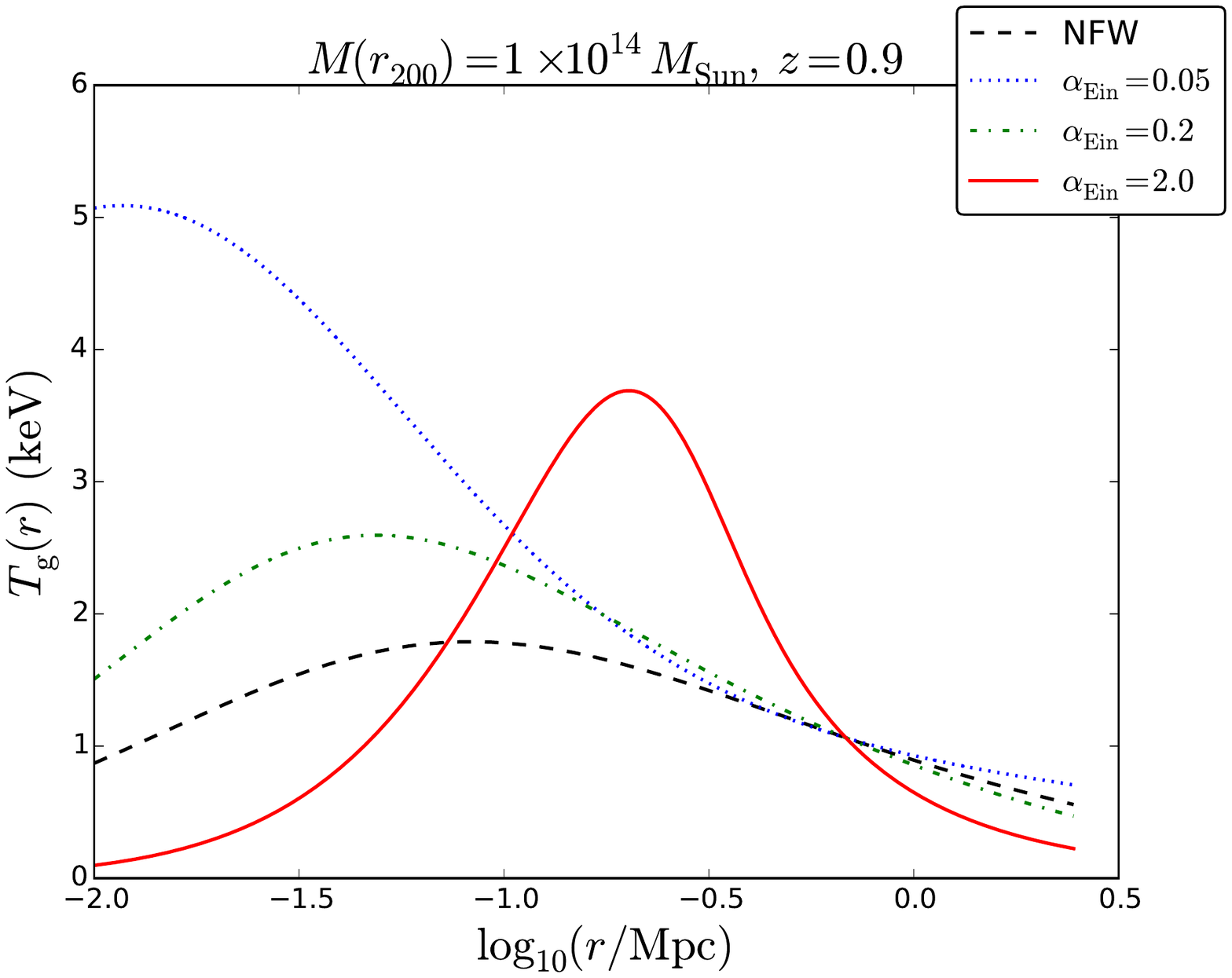} &
     \includegraphics[ width=0.5\linewidth]{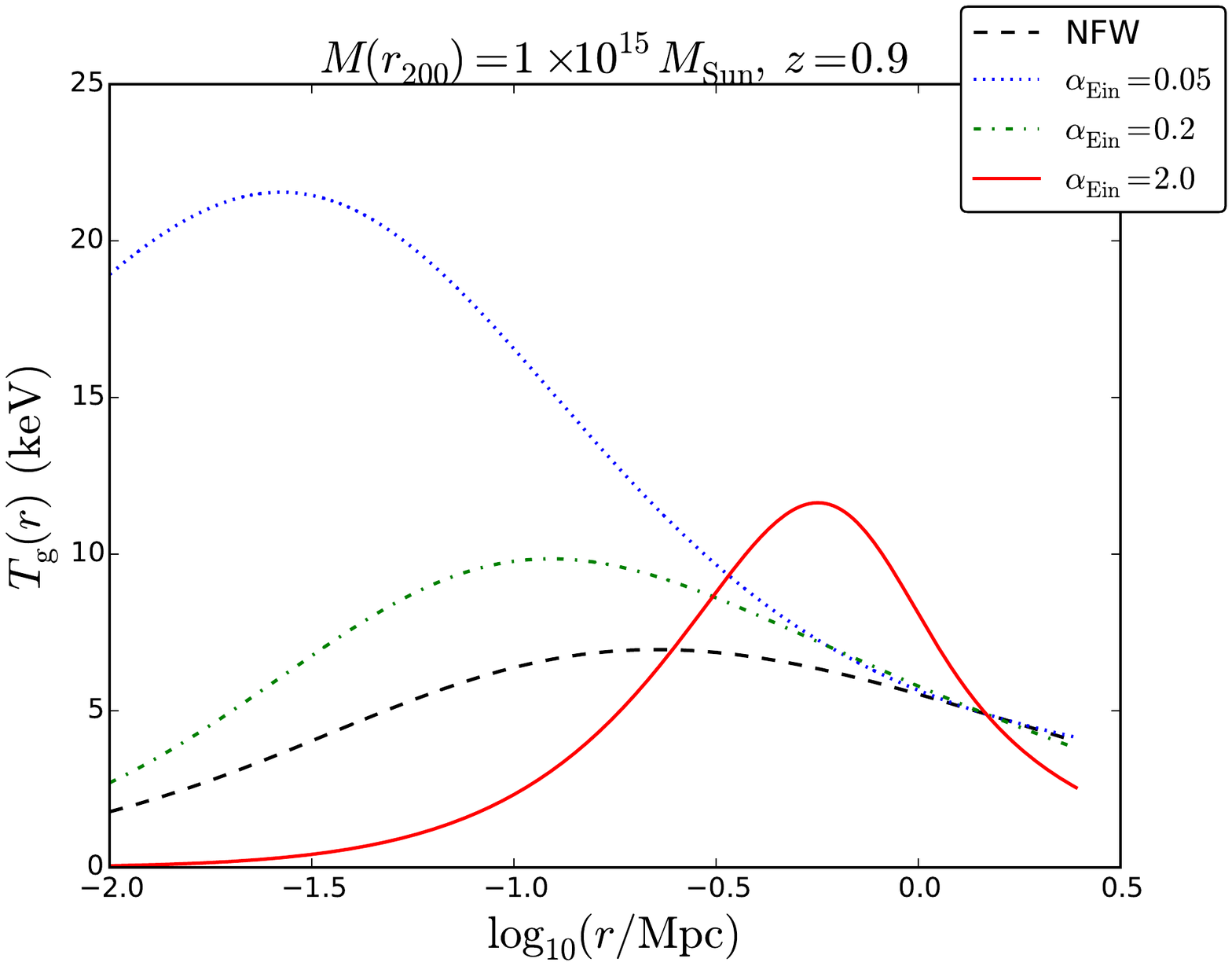} \\
    \end{tabular}
  \caption{Gas temperature profiles as a function of log cluster radius using NFW and Einasto models. Values of $\alpha_{\mathrm{Ein}} = 0.05$, $0.2$, and $2.0$ are used as inputs. Top row has $z = 0.15$, bottom row has $z = 0.9$. Left column has $M(r_{200}) = 1\times 10^{14} M_{\mathrm{Sun}}$, right column has $M(r_{200}) = 1\times 10^{15} M_{\mathrm{Sun}}$.}
\label{graph:einnfwgt}
  \end{center}
\end{figure*}

Gas temperature profiles are shown in Figure~\ref{graph:einnfwgt}. The $\alpha_{\rm Ein} = 2.0$ is very distinctive, always peaking at much higher $r$ than the other three and also always much more sharply.

\subsection{Bayesian analysis of AMI data}
\label{subsec:results2}

We now focus our attention on applying the PM II to simulated and real AMI data, to compare the parameter estimates and Bayesian evidences with those obtained from the PM I.


\subsubsection{Simulated AMI data}
\label{subsubsec:simulated}
\citet{2016JCAP...01..042S} study the errors associated with fitting NFW profiles to Einasto dark matter halos and vice versa for weak lensing studies. We conduct similar work in the context of simulated SZ observations. The simulations were carried out using the in-house AMI simulation package \textsc{Profile}, which has been used in various forms in e.g. \citet{2002MNRAS.333..318G}, \citet{2013MNRAS.430.1344O} and KJ19. \\
As before we consider Einasto profiles with the $\alpha_{\rm Ein}$ values $0.05$, $0.2$, and $2.0$ plus an NFW profile. each with $M(r_{200}) = 1\times 10^{14} M_{\mathrm{Sun}}$ or $M(r_{200}) = 1\times 10^{15} M_{\mathrm{Sun}}$, $z = 0.15$ or $z = 0.90$ and $f_{\rm gas}(r_{200}) = 0.12$. Note for all of these simulations no radio-sources, primordial CMB or confusion noise were included, and instrumental noise was set to a negligible level.

We first compare the posterior distributions for the input parameters (except those with $\delta$-function priors). The posterior distributions are plotted using \textsc{GetDist}\footnote{\url{http://getdist.readthedocs.io/en/latest/}.}, and the contours on the two-dimensional plots represent the 95\% and 68\% mean confidence intervals.
Table~\ref{tab:simrestab} in Appendix~\ref{sec:simrestab} summarises the input and output values of the 16 simulations. The output values are the marginalised posterior mass mean estimates and standard deviations. The first column gives the model used to \textit{simulate} the cluster, with the following two columns giving the mass and $z$ input values. For each simulation, we \textit{analysed} the data using two models, one using the NFW profile and one using an Einasto profile. For data simulated using an NFW profile, when analysing the data with an Einasto profile we used $\alpha_{\rm Ein} = 0.2$. For data simulated using an Einasto profile, when analysing the data with an Einasto profile we set $\alpha_{\rm Ein}$ equal to the value used as the input for the simulation. We also repeated each simulation 10 times with different noise realisations to check the statistical significance of our results.

\begin{figure}
  \begin{center}
  \includegraphics[ width=\linewidth]{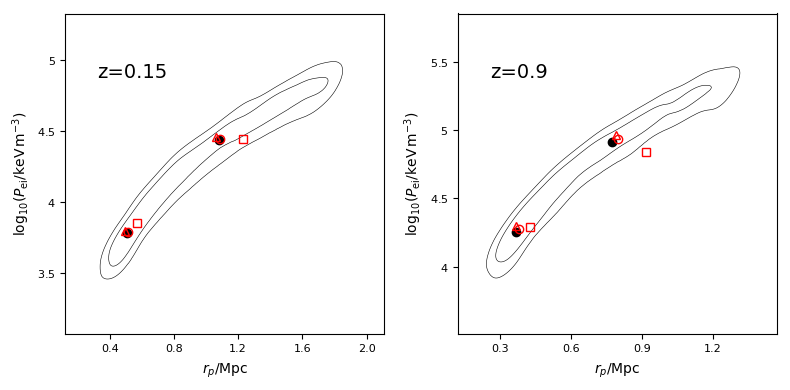}
  \caption{Prior induced on the $r_{\rm p}$ and $P_{\rm ei}$ parameters describing the gas pressure profile using PM~I (black contours; 68 and 95\% levels), for the two simulation redshifts, overplotted with the true values for our simulation set.  Black dots are PM~I; red empty triangles, circles and squares are PM~II with $\alpha_{\rm Ein} = 0.05$, 0.2, 2.0 respectively.  The set of points with lower $r_{\rm p}$/$P_{\rm ei}$ values correspond to the $M(r_{200})=1\times 10^{14} M_{\mathrm{Sun}}$ simulations and the set with higher values to the $M(r_{200})=10\times 10^{14} M_{\mathrm{Sun}}$ simulations.}
\label{Fi:rp_Pei_comp}
  \end{center}
\end{figure}

We firstly note that the Bayesian evidence values for the Einasto and NFW analyses are the same within the errors in almost all cases, with only the $\alpha_{\rm Ein} = 2.0$ and $M(r_{200}) = 10 \times 10^{14} M_{\rm Sun}$ cases showing a weak preference for the correct model.  This can be explained as follows.  Both models implement a GNFW profile for the pressure distribution, with the physical model providing the characteristic scale and normalisation parameters $r_{\rm p}$ and $P_{\rm ei}$ (see equation~\ref{eqn:epressure}).  Therefore the SZ signal is entirely described by these two parameters, and provided that the correct pair of values can be reached using the physical model used for the analysis, the SZ signal can be described equally well by either model.  This is illustrated in Figure~\ref{Fi:rp_Pei_comp}, where we show the prior on $r_{\rm p}$ and $P_{\rm ei}$ induced by PM~I for the two redshifts of our simulations, over-plotted with the true values for each of the simulations.  It can be seen that the $\alpha_{\rm Ein} = 2.0$ and $M(r_{200}) = 10 \times 10^{14} M_{\rm Sun}$ cases are the only ones for which the true value is significantly outside the prior, explaining the reduced evidence values.

\begin{figure}
  \begin{center}
  \includegraphics[ width=\linewidth]{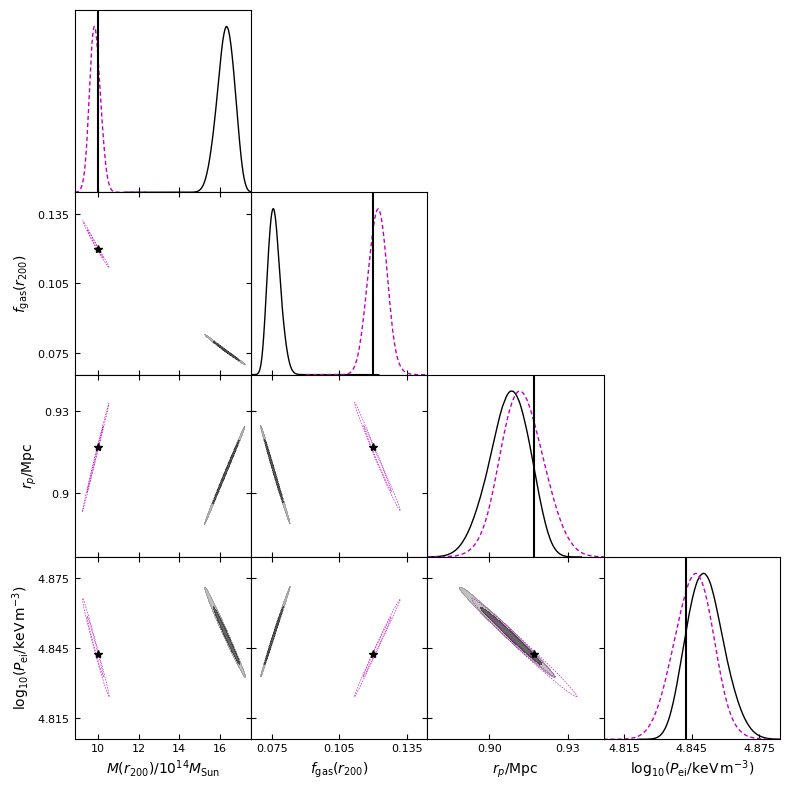}
  \caption{Posterior distributions for cluster simulated with $\alpha_{\rm Ein} = 2.0$, $M(r_{200}) = 1\times 10^{15} M_{\mathrm{Sun}}$ and $z=0.9$, modelled with: Einasto dark matter profile (black filled contours and solid lines), and NFW dark matter profile (magenta empty contours and dashed lines).  True values are shown with black stars and vertical lines.}
\label{Fi:a2.0_M10_z0.9_post}
  \end{center}
\end{figure}

However, despite the identical evidence values, the mass constraints are clearly biased when using the wrong model as the mapping from $M(r_{200})$ and $f_{\rm gas}(r_{200})$ to $r_{\rm p}$ and $P_{\rm ei}$ differs depending on the model.  When simulating and analysing with the same model, we find the mean of the mass posterior is within $1\sigma$ of the input value in 14 out of 16 cases (82\% over all the noise realisations); this is higher than the expected 68\% due to the additional information provided by the prior.  When analysing the simulation with the wrong model, we find this in only 6 out of 16 cases (42\% over all the noise realisations).  In Figure~\ref{Fi:a2.0_M10_z0.9_post} we show the posteriors produced when analysing the PM~II simulation with $\alpha_{\rm Ein} = 2.0$, $M(r_{200}) = 10 \times 10^{14} M_{\rm Sun}$, $z=0.9$ with PM~II and PM~I.  The correct $r_{\rm p}$ and $P_{\rm ei}$ values are recovered in both cases, however the mass and gas fraction posteriors are strongly biased from their true values when analysing with PM~I.  The very low $f_{\rm gas}(r_{200})$ value, far outside the prior, can be understood by considering Figure~\ref{Fi:rp_Pei_comp}; the $r_{\rm p}$/$P_{\rm ei}$ prior is `thickened' by allowing a greater range in $f_{\rm gas}(r_{200})$, so to reach the $r_{\rm p}$/$P_{\rm ei}$ parameter pair outside the prior, $f_{\rm gas}(r_{200})$ must be dragged outside its prior range.

Finally, we tried running the Bayesian analysis on eight of the Einasto simulated clusters with uniform analysis priors on $\alpha_{\rm Ein}$. These clusters corresponded to the simulations with input values of either $\alpha_{\rm Ein} = 0.2$ or $\alpha_{\rm Ein} = 2.0$. 

\begin{figure*}
  \begin{center}
  \includegraphics[ width=0.33\linewidth]{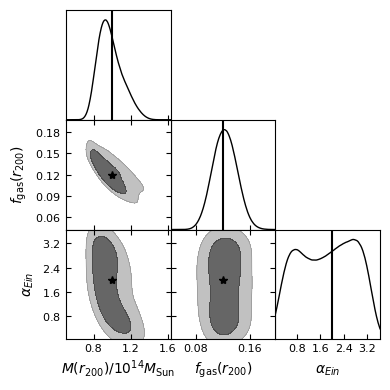}
  \includegraphics[ width=0.33\linewidth]{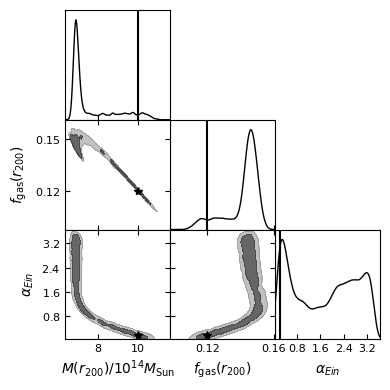}
  \includegraphics[ width=0.33\linewidth]{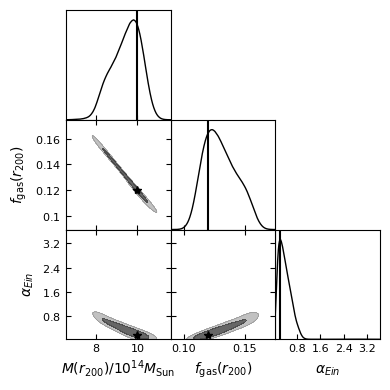}
  \medskip
  \centerline{(a) \hskip 0.33\linewidth (b) \hskip 0.33\linewidth (c)}
  \caption{Posterior distributions of Einasto model input parameters for: (a) $\alpha_{\rm Ein} = 2.0$, $M(r_{200}) = 1\times 10^{14} M_{\mathrm{Sun}}$ and $z=0.9$, (b) $\alpha_{\rm Ein} = 0.2$, $M(r_{200}) = 10\times 10^{14} M_{\mathrm{Sun}}$ and $z=0.15$ simulated cluster, and (c) $\alpha_{\rm Ein} = 0.2$, $M(r_{200}) = 10\times 10^{14} M_{\mathrm{Sun}}$ and $z=0.9$ simulated clusters.}
\label{graph:eingoodalpha}
  \end{center}
\end{figure*}

\begin{figure}
  \begin{center}
  \includegraphics[ width=\linewidth]{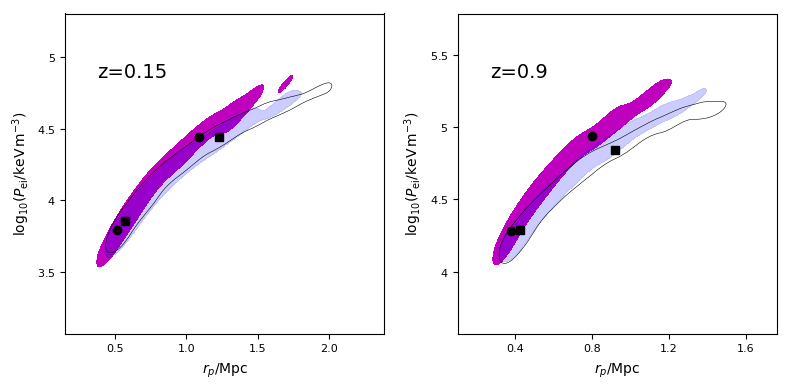}
  \caption{Prior induced on the $r_{\rm p}$ and $P_{\rm ei}$ parameters describing the gas pressure profile using PM~II, for a representative set of $\alpha_{\rm Ein}$ values, 0.2 (solid magenta), 1.0 (transparent blue) and 2.0 (empty black), where the contours shown are the 68\% levels, for the two simulation redshifts.  The true values for the simulations analysed varying $\alpha_{\rm Ein}$ are overplotted with circles ($\alpha_{\rm Ein}=0.2$) and squares ($\alpha_{\rm Ein}=2.0$).  The set of points with lower $r_{\rm p}$/$P_{\rm ei}$ values correspond to the $M(r_{200})=1\times 10^{14} M_{\mathrm{Sun}}$ simulations and the set with higher values to the $M(r_{200})=10\times 10^{14} M_{\mathrm{Sun}}$ simulations.}
\label{Fi:rp_Pei_comp2}
  \end{center}
\end{figure}

In both cases we analysed the simulations with a uniform prior on $\alpha_{\rm Ein}$, $\mathcal{U}[0.05, 3.5]$.  The results fell into three general categories, examples of which are shown in Figure~\ref{graph:eingoodalpha}.  In the first category, $\alpha_{\rm Ein}$ was completely unconstrained and the mass posterior was just widened by the marginalization over $\alpha_{\rm Ein}$.  This was generally the case for the lower-mass clusters with $\alpha_{\rm Ein}=2.0$, which falls more toward the centre of the allowed prior range (Figure~\ref{graph:eingoodalpha}a).  In most other cases, large, curving degeneracies were seen between the parameters which induced biases in the one-dimensional marginalised mass constraints, although the true value was correctly located within the two-dimensional constraints; a particularly severe example of this is shown in Figure~\ref{graph:eingoodalpha}b.  In only one case, the cluster with $\alpha_{\rm Ein} = 0.2$, $M(r_{200}) = 10\times 10^{14} M_{\mathrm{Sun}}$ and $z=0.9$, a strong constraint with little degeneracy was produced on all parameters (Figure~\ref{graph:eingoodalpha}c).  These results can again be understood by considering the priors induced by the physical model with a given value of $\alpha_{\rm Ein}$ on $r_{\rm p}$ and $P_{\rm ei}$, as shown in Figure~\ref{Fi:rp_Pei_comp2}.  The constraint on $\alpha_{\rm Ein}$ depends entirely on whether the ($r_{\rm p}$, $P_{\rm ei}$) parameter pair can be reached using PM~II with a given $\alpha_{\rm Ein}$ value; the case of $\alpha_{\rm Ein} = 0.2$, $M(r_{200}) = 10\times 10^{14} M_{\mathrm{Sun}}$ and $z=0.9$ is the only one where the priors induced by the different physical models are quite separate and therefore the only one with a strong constraint.  The difference between the priors on $r_{\rm p}$ and $P_{\rm ei}$ at $z=0.15$ and $z=0.9$ is simply produced by the differences in $c_{200}$ and $\rho_{\rm crit}$ at the different redshifts.

We therefore see that although $\alpha_{\rm Ein}$ can be constrained using SZ data in some cases, in general it is unconstrained and varying it can introduce biases in the one-dimensional marginalised mass estimates.  Referring back to the finding by \citet{2016MNRAS.457.4340K} that $\alpha_{\rm Ein}$ and cluster mass are positively correlated, it makes sense in future work to either incorporate a joint prior on $M$ and $\alpha_{\rm Ein}$, or to include a functional form between the two variables in the cluster models.

\subsubsection{Analysis of real AMI observations of A611}
\label{subsubsec:a611}

\begin{figure}
  \begin{center}
  \includegraphics[ width=0.90\linewidth]{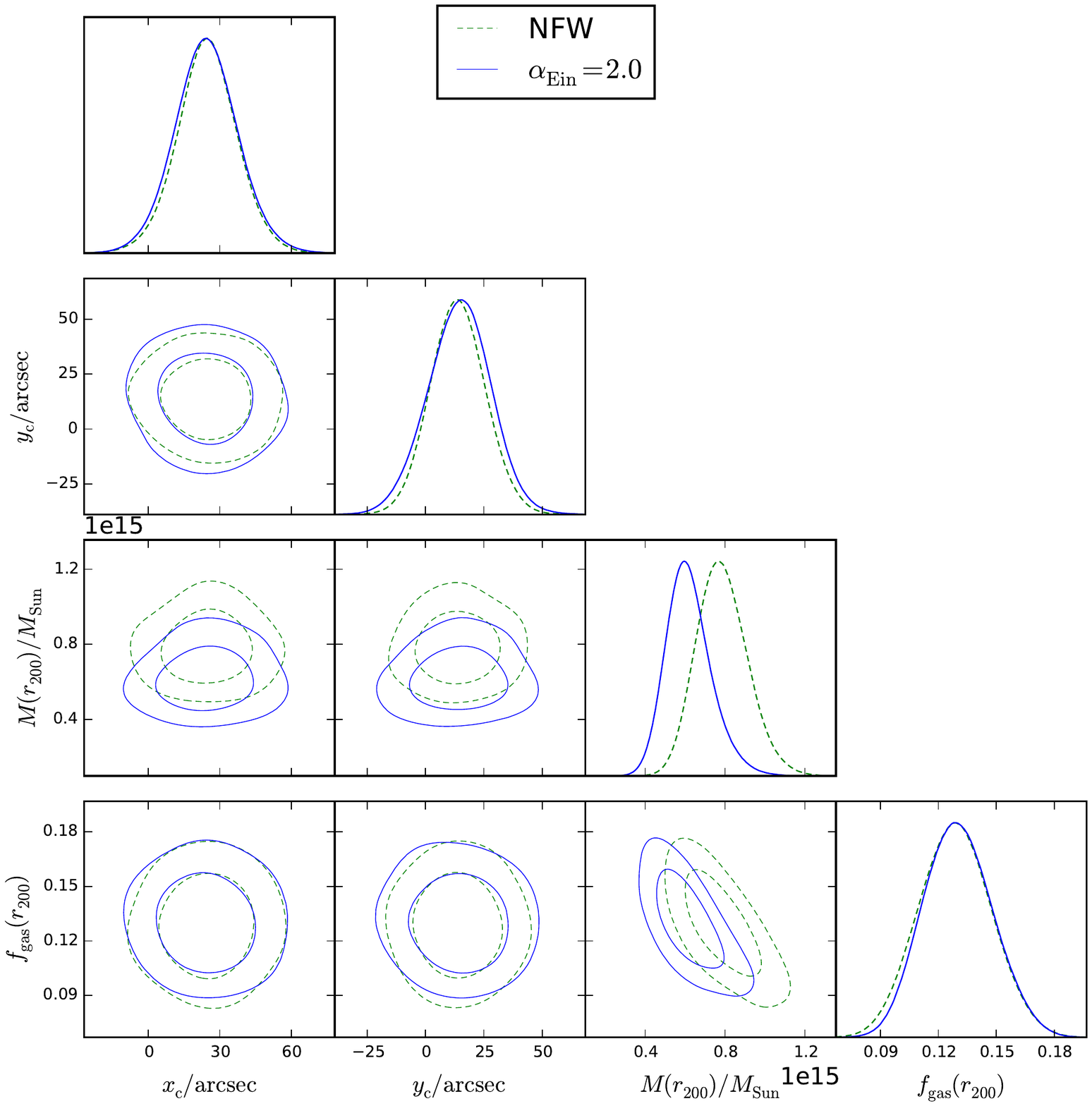}
  \caption{Marginalised posterior distributions of physical model input parameters for the NFW and $\alpha_{\rm Ein} = 2.0$ models applied to real A611 data. The contour plots are the two-dimensional marginalised plots of the parameters named in the corresponding row / column. The line plots are the fully marginalised posterior distributions.}
\label{graph:a611posterior}
  \end{center}
\end{figure}

Following MO12 we conduct Bayesian analysis on data from observations with AMI of the cluster A611 at $z = 0.288$, which has been studied through its X-ray emission, strong lensing, weak lensing and SZ effect (see \citealt{2007MNRAS.379..209S}, \citealt{2011A&A...528A..73D}, \citealt{2010A&A...514A..88R} and \citealt{2016MNRAS.460..569R} respectively). These studies suggest that there is no significant contamination from radio-sources and that the cluster has similar weak lensing and X-ray masses and is close to the $T_{\rm{X}}$--$T_{\rm{SZ}}$ relation for clusters close to hydrostatic equilibrium. \\
Referring back to Section~\ref{subsubsec:priors}, we take $z = 0.288$ and for the analysis with the PM II we consider three different cases separately -- $\alpha_{\rm Ein} \in \{0.05, 0.2, 2.0\}$ -- so that in total there are four sets of results to compare for A611 (including the NFW model). The Bayesian analysis was conducted in the same way as it was for the simulations in the previous Section. \\
The means and standard deviations of the four analyses are given in Table~\ref{tab:a611results}. As in Section~\ref{subsec:results1}, $\alpha_{\rm Ein} = 0.05$ and $\alpha_{\rm Ein} = 0.2$ show similar results to PM I. $\alpha_{\rm Ein} = 2$ gives a different estimate for $M(r_{200})$, and its posterior distribution is shown in Figure~\ref{graph:a611posterior} along with that obtained with the NFW profile. The mean mass estimates are within one combined standard deviation away from each other. However, as seen in Table~\ref{tab:a611results} the value of $\ln(\mathcal{Z}_{\rm{Ein}} / \mathcal{Z}_{\rm NFW})$ imply that `no model is favoured by the data' according to the Jeffreys scale. This is entirely consistent with the results obtained from the simulations in the previous Section.

\begin{table*}
\begin{center}
\begin{tabular}{{l}{c}{c}{c}{c}{c}}
\hline
Model & $x_{\rm c}$~(arcsec) & $y_{\rm c}$~(arcsec) & $M(r_{200})$~($\times 10^{14}M_{\mathrm{Sun}}$) & $f_{\rm gas}(r_{200})$ & $\ln \left(\mathcal{Z}\right)$ \\ 
\hline
NFW & $24.7 \pm 12.4$ & $13.9 \pm 11.5$ & $7.84 \pm 1.24$ & $0.129 \pm 0.020$ & $38629.4 \pm 0.3$ \\
$\alpha_{\rm Ein} = 0.05$  & $22.7 \pm 12.5$ & $13.1 \pm 12.6$ & $7.45 \pm 1.24$ & $0.130 \pm 0.019$ & $38629.2  \pm 0.3$ \\
$\alpha_{\rm Ein} = 0.2$  & $25.5 \pm 12.8$ & $14.9 \pm 13.0$ & $7.67 \pm 1.27$ & $0.127 \pm 0.017$ & $38629.7  \pm 0.2$ \\
$\alpha_{\rm Ein} = 2.0$  & $24.3 \pm 12.4$ & $14.3 \pm 13.2$ & $6.17 \pm 1.12$ & $0.130 \pm 0.017$ & $38629.2 \pm 0.2$ \\
\hline
\end{tabular}
\caption{Marginalised posterior distribution mean values and standard deviations of physical model input parameters and Bayesian evidences associated with each model, applied to real A611 data.}
\label{tab:a611results}
\end{center}
\end{table*}

\section{Forecasting}\label{sec:forecasting}

Our simulations have shown that SZ data are generally unable to distinguish between these physical models based on the thermal SZ effect alone; our simulations have a thermal noise level $\approx$\,$100\times$ smaller than a typical AMI observation and include no primordial CMB or radio source confusion noise (or other instrumental systematics).  However, recently it has been shown that \emph{Planck} cluster constraints may be biased by the relativistic SZ corrections which depend on the temperature of the cluster (e.g.\ \citealt{2018MNRAS.476.3360E}), and including these corrections will become crucial for forthcoming instruments with higher sensitivity and angular resolution such as CCAT-prime (e.g.\ \citealt{2018SPIE10700E..5XP}).  Our models provide a consistent physical mechanism for modelling and including the relativistic corrections. It has been shown in \citet{2018JCAP...02..032M} that assuming isothermality biases cluster constraints based on simulated CCAT-prime data, so given the very different temperature profiles produced by our models (see Figure~\ref{graph:einnfwgt}), forthcoming instruments may be able to distinguish between these models based on spatially-resolved relativistic SZ constraints, although the effects of cooling flow and merger activity must also be considered.


\section{Conclusions}
\label{sec:summary}
Based on the physical model introduced in \citet{2012MNRAS.423.1534O} (PM I) which uses an NFW profile \citep{1995MNRAS.275..720N} to model the dark matter content of galaxy clusters, we derive a new physical model (PM II) which models the dark matter with an Einasto profile \citep{1965TrAlm...5...87E}.
The Einasto profile has an additional degree of freedom compared to the NFW profile, which dictates the shape of the dark matter density as a function of radius. For different values of $\alpha_{\rm Ein}$ we have investigated the profiles of several physical properties of a cluster, namely the dark matter density, dark matter mass, gas density, gas mass and gas temperature. We have also provided the equivalent profiles in the NFW case. From this we found the following.
\begin{itemize}
\item Of the three values of $\alpha_{\rm Ein}$ considered, $\alpha_{\rm Ein} = 0.2$ gave the most similar profile to that given by the NFW model (as discussed in \citealt{2014MNRAS.441.3359D}), with the main discrepancy between the two arising in the peak amplitude of the gas temperature.
\item $\alpha_{\rm Ein} = 2.0$ showed the most convergent behaviour in $M_{\rm dm}(r)$ at high $r$, but the most divergent in $M_{\rm g}(r)$ in the same limit.
\item The gas temperature profiles were somewhat different for the $\alpha_{\rm Ein}$ values considered here. This suggests that if one can carefully measure the temperature profile of a cluster, then one could infer $\alpha_{\rm Ein}$ and use this in the model presented here (though one has to be aware of cooling flow and merger activity).
\end{itemize}
Next we applied Bayesian analysis to simulated and AMI datasets using PM I and PM II, to compare the models' parameter estimates and fits to the data. 
Simulating clusters with either NFW or Einasto dark matter profiles, which were then `observed' by AMI, we found the following.
\begin{itemize}
\item When the wrong cluster model is used in the analysis, the correct mass value is inferred to within $1\sigma$ for only 6 of the 16 clusters (compared to 14 out of 16 for the correct model). In certain cases, the gas mass fraction at $r_{200}$ is also inferred incorrectly, which in turn is due to the true values of the GNFW scale and normalisation parameters ($r_{\rm p}$ and $P_{\rm ei}$) laying outside of the model's priors for these parameters.
\item Looking at the Bayesian evidence values for the simulations, for 14 of the 16 clusters no model is preferred over the other one according to the Jeffreys scale.
\item The two simulations which did show preference towards a model picked the correct one, with `weak preference' according to the Jeffreys scale. Both of these simulations were generated with PM~II (and showed preference to PM~II over PM~I). From inspection of the priors on $r_{\rm p}$ and $P_{\rm ei}$ for PM~I for these two clusters, it was apparent that the true value was far away from the prior peaks, thus explaining the low evidence values for these runs.
\item When $\alpha_{\rm Ein}$ was allowed to vary in the analysis, it was found to be mostly unconstrained (except in one exceptional case), and the large, curving degeneracies between $\alpha_{\rm Ein}$ and the other cluster parameters could produce biases in the one-dimensional mass constraints.  To safely allow variation in $\alpha_{\rm Ein}$, a physically-motivated relationship between $M$ and $\alpha_{\rm Ein}$ could be introduced such as that found by \citet{2016MNRAS.457.4340K}.
\end{itemize}
Using real data from cluster A611 we found, consistent with the simulations:
\begin{itemize}
\item The $\alpha_{\rm Ein} = 0.05$ and $\alpha_{\rm Ein} = 0.2$ models gave very similar results to the NFW model; the $\alpha_{\rm Ein} = 2$ model however underestimates $M(r_{200})$ relative to the other three models.
\item The Bayesian evidence values calculated from these four analyses were roughly equal, suggesting no model provided a statistically more significant fit relative to the others.
\end{itemize}
In a forthcoming paper \citet{2019MNRAS.486.2116P}, Bayesian analysis will be performed on joint AMI-\textit{Planck} datasets.


\section*{Acknowledgements}
We are grateful to the anonymous referee for a wealth of well-judged, constructive and thorough comments that have much improved this paper.
This work was performed using the Darwin Supercomputer of the University of Cambridge High Performance Computing (HPC) Service (\url{http://www.hpc.cam.ac.uk/}), provided by Dell Inc. using Strategic Research Infrastructure Funding from the Higher Education Funding Council for England and funding from the Science and Technology Facilities Council. The authors would like to thank Stuart Rankin from HPC and Greg Willatt and David Titterington from Cavendish Astrophysics for computing assistance. They would also like to thank Dave Green for his invaluable help using \LaTeX.
Kamran Javid acknowledges an STFC studentship. Yvette Perrott acknowledges support from a Trinity College Junior Research Fellowship and Rutherford Discovery Fellowship.


\setlength{\bibsep}{0pt}            
\renewcommand{\bibname}{References} 


\appendix
\section{Einasto mass integral}
\label{sec:einastointegral}

From equations~\ref{eqn:einasto} and~\ref{eqn:massrho} we have that
\begin{equation}
\label{eqn:appendixmassrho}
\begin{split}
M(r) &= \int_{0}^{r} 4\pi r'^{2} \rho_{\rm dm, PM II}(r') \, \rm{d}r' \\
     &= 4 \pi \rho_{-2} \exp \left( 2 / \alpha_{\rm Ein} \right) \int_{0}^{r} r'^{2} \exp \left[ \frac{-2}{\alpha_{\rm Ein}} \left( \frac{r'}{r_{-2}} \right)^{\alpha_{\rm Ein}} \right] \rm{d} r'.
\end{split}
\end{equation}
Using the substitution 
\begin{equation}
\label{eqn:appendixeinsub1}
u = \frac{2^{3/ \alpha_{\rm Ein}} r'^{3}}{\alpha_{\rm Ein}^{3 / \alpha_{\rm Ein}} r_{-2}^{3}} \Rightarrow \rm{d} u = \frac{3 \times 2^{3 / \alpha_{\rm Ein}}r'^{2}}{ \alpha_{\rm Ein} ^{3 / \alpha_{\rm Ein}}r_{-2}^{3}} \rm{d}r'
\end{equation}
then equation~\ref{eqn:appendixmassrho} becomes
\begin{equation}
\label{eqn:appendixmassrho2}
\begin{split}
M(r) &= \frac{4 \pi \rho_{-2} \exp \left( 2 / \alpha_{\rm Ein} \right) \alpha_{\rm Ein}^{3 / \alpha_{\rm Ein}} r_{-2}^{3}}{3 \times 2^{3 / \alpha_{\rm Ein}}} \\
       & \quad \times \int_{u = 0}^{u = \frac{2^{3/ \alpha_{\rm Ein}} r^{3}}{\alpha_{\rm Ein}^{3 / \alpha_{\rm Ein}} r_{-2}^{3}}} \exp \left( -u^{\alpha_{\rm Ein} / 3} \right) \rm{d} u.
\end{split}
\end{equation}
Finally, using the substitution $t = u^{\alpha_{\rm Ein} / 3}$ so that $\rm{d}t = \frac{\alpha_{\rm Ein}}{3} u^{\alpha_{\rm Ein} / 3 -1} \rm{d}u$, then the integral in equation~\ref{eqn:appendixmassrho2} (ignoring the constant factor) becomes
\begin{equation}
\label{eqn:appendixmassrho3}
\begin{split}
\frac{3}{\alpha_{\rm Ein}} \int_{0}^{u^{\alpha_{\rm Ein}/3}} u^{1 - \alpha_{\rm Ein} / 3} e^{-t} \rm{d}t &= \frac{3}{\alpha_{\rm Ein}} \int_{0}^{\frac{2r^{\alpha_{\rm Ein}}}{\alpha_{\rm Ein}r_{-2}^{\alpha_{\rm Ein}}}} t^{ 3 / \alpha_{\rm Ein} - 1} e^{-t} \rm{d}t \\
                                                                                                         &= \gamma \left[ \frac{3}{\alpha_{\rm Ein}}, \frac{2}{\alpha_{\rm Ein}} \left( \frac{r}{r_{-2}} \right)^{\alpha_{\rm Ein}} \right],
\end{split}
\end{equation} 
where the last equality follows from the definition of the incomplete lower Gamma function $\gamma \left[a, x \right] = \int_{0}^{x} t^{a-1} e^{-t} \rm{d} t $. Including the constant factor in equation~\ref{eqn:appendixmassrho2} leads to the result
\begin{equation}
\label{eqn:appendixmassrho4}
M(r) = \frac{4 \pi \rho_{-2}}{\alpha_{\rm Ein}} \exp( 2 / \alpha_{\rm Ein}) \left( \frac{\alpha_{\rm Ein}}{2} \right) ^{3 / \alpha_{\rm Ein}} \gamma \left[ \frac{3}{\alpha_{\rm Ein}}, \frac{2}{\alpha_{\rm Ein}} \left( \frac{r}{r_{-2}} \right)^{\alpha_{\rm Ein}} \right].
\end{equation}


\section{Determining $\lowercase{r}_{500}$ iteratively}
\label{sec:r500newton}

Evaluating equations~\ref{eqn:massrhocrit} and~\ref{eqn:massrho} at $r_{500}$ and equating we get 
\begin{equation}
\label{eqn:r500iter1}
\begin{split}
\frac{4\pi}{3} 500 \rho_{\rm crit} (z) r_{500}^{3} &= 4 \pi \rho_{-2} 1 / \alpha_{\rm Ein} \exp( 2 / \alpha_{\rm Ein}) \left( \frac{\alpha_{\rm Ein}}{2} \right) ^{3 / \alpha_{\rm Ein}} r_{-2}^{3} \\
                                                   & \quad \times \gamma \left[ \frac{3}{\alpha_{\rm Ein}}, \frac{2}{\alpha_{\rm Ein}} \left( \frac{r_{500}}{r_{-2}} \right)^{\alpha_{\rm Ein}} \right].
\end{split}
\end{equation}
If we let $R = r_{500} / r_{-2}$, then we can determine $r_{500}$ by solving the following for $R$
\begin{equation}
\label{eqn:r500iter2}
\begin{split}
&\frac{R^{3}}{\gamma \left[ \frac{3}{\alpha_{\rm Ein}}, \frac{2}{\alpha_{\rm Ein}} R^{\alpha_{\rm Ein}} \right]} \\ 
& - \frac{1}{\rho_{\rm crit}(z)} \frac{3 \rho_{-2}}{500} \left(\frac{\alpha_{\rm Ein}}{2}\right)^{3 / \alpha_{\rm Ein}} \frac{\exp \left( 2 / \alpha_{\rm Ein} \right)}{\alpha_{\rm Ein}} = 0
\end{split}
\end{equation}
by some iterative root-finding method e.g. Newton-Raphson. We use the starting point $R_{0} = \frac{2r_{200}}{3r_{-2}}$ which usually results in the algorithm converging in $\mathcal{O}(10)$ iterations. \\

We now show that equation~\ref{eqn:r500iter2} only has one solution for a given $r_{-2}$. We start by considering both sides of equation~\ref{eqn:r500iter1} as two different functions, and ignore constant terms for simplicity (this does not affect the truth of the final result), i.e. we consider the two functions
\begin{equation}
\label{eqn:r500iter3}
f(r_{500}) = r_{500}^{3}, \, g(r_{500}) = \gamma \left[ \frac{3}{\alpha_{\rm Ein}}, \frac{2}{\alpha_{\rm Ein}} \left( \frac{r_{500}}{r_{-2}} \right)^{\alpha_{\rm Ein}} \right].
\end{equation}
We first note that $f(0) = g(0) = 0$, and differentiate both functions with respect to $r_{500}$
\begin{equation}
\label{eqn:r500iter4}
\frac{\mathrm{d}f}{\mathrm{d}r_{500}} \propto r_{500}^{2}, \,  \frac{\mathrm{d}g}{\mathrm{d}r_{500}} \propto r_{500}^{2} \exp \left[ -\frac{2}{\alpha_{\rm Ein}} \left( \left(\frac{r_{500}}{r_{-2}}\right)^{\alpha_{\rm Ein}} - 1 \right) \right].
\end{equation}
Setting these two derivatives equal to each other yields one solution at $r_{500}=r_{-2}$ for all $\alpha_{\rm Ein} \neq 0$, meaning the derivatives only intersect once. Furthermore $\frac{\mathrm{d}g}{\mathrm{d}r_{500}}$ tends to zero for large $r_{500}$ whilst $\frac{\mathrm{d}f}{\mathrm{d}r_{500}}$ is a monotonically increasing function, meaning the former must be larger before the two intersect. This coupled with the fact that $f(0) = g(0) = 0$ means that $g(r_{500}) > f(r_{500})$ until some point (which has to be after the derivatives intersect) when the two intersect, after which $f(r_{500}) > g(r_{500})$ as $g(r_{500})$ flattens off.
This proves that equation~\ref{eqn:r500iter2} only has one root and that equation~\ref{eqn:r500iter1} only has one solution in $r_{500}$ for fixed $r_{-2}$.


\newpage
\newgeometry{margin=1cm} 
\onecolumn
\begin{landscape}
\section{Simulation results table}
\label{sec:simrestab}
\begin{center}
\begin{longtable}{llllllll}
\caption{Input and output values of simulations using NFW and Einasto dark matter profiles. The first column is what dark matter profile was used to \textit{simulate} the cluster. Input $M(r_{200})$ and Input $z$ are the input values used to create the simulation for the given model. Ein out $M(r_{200})$ is the mean and standard deviation of the posterior distribution obtained inferred using an Einasto profile to \textit{model} the cluster. Ein $\ln(\mathcal{Z})$ is the natural log Bayesian evidence corresponding to the inference. NFW... is as before but using an NFW profile in the \textit{modelling}. $\ln(\mathcal{Z}_{\rm Ein} / \mathcal{Z}_{\rm NFW}) $ is the natural log ratio of the two evidences obtained.}
\label{tab:simrestab} \\

\hline \multicolumn{1}{c}{Model} & \multicolumn{1}{c}{Input $M(r_{200})~(\times10^{14}M_{\mathrm{Sun}})$} & \multicolumn{1}{c}{Input $z$} & \multicolumn{1}{c}{Ein out $M(r_{200})~(\times10^{14}M_{\mathrm{Sun}})$} & \multicolumn{1}{c}{NFW out $M(r_{200})~(\times10^{14}M_{\mathrm{Sun}})$} & \multicolumn{1}{c}{Ein $\ln(\mathcal{Z})$} & \multicolumn{1}{c}{NFW $\ln(\mathcal{Z})$} & \multicolumn{1}{c}{$\ln(\mathcal{Z}_{\rm Ein} / \mathcal{Z}_{\rm NFW}) $}  \\ \hline 

\tabulinesep=_1mm
\extrarowsep=1mm
\LTcapwidth=\textwidth

$\alpha_{\rm Ein} = 0.2$ & 1 & 0.15 & $0.95 \pm 0.10$ & $0.96 \pm 0.10$ & $30061.7 \pm 0.2$ & $30061.8 \pm 0.2$ & $-0.1 \pm 0.2$ \\
$\alpha_{\rm Ein} = 2.0$ & 1 & 0.15 & $1.07 \pm 0.12$ & $1.46 \pm 0.15$ & $30114.8 \pm 0.2$ & $30114.4 \pm 0.2$ & $0.5 \pm 0.2$ \\
$\alpha_{\rm Ein} = 0.05$ & 1 & 0.15 & $0.95 \pm 0.10$ & $0.94 \pm 0.11$ & $30034.6 \pm 0.2$ & $30034.6 \pm 0.2$ & $0.0 \pm 0.2$ \\
NFW & 1 & 0.15 & $1.02 \pm 0.12$ & $1.03 \pm 0.09$ & $29991.3 \pm 0.2$ & $29991.1 \pm 0.2$ & $0.2 \pm 0.2$ \\
$\alpha_{\rm Ein} = 0.2$ & 1 & 0.90 & $0.99 \pm 0.12$ & $1.08 \pm 0.13$ & $30059.9 \pm 0.1$ & $30059.8 \pm 0.1$ & $0.1 \pm 0.2$ \\
$\alpha_{\rm Ein} = 2.0$ & 1 & 0.90 & $0.95 \pm 0.12$ & $1.27 \pm 0.13$ & $30098.9 \pm 0.2$ & $30098.8 \pm 0.2$ & $0.1 \pm 0.2$ \\
$\alpha_{\rm Ein} = 0.05$ & 1 & 0.90 & $1.13 \pm 0.12$ & $1.22 \pm 0.15$ & $29967.8 \pm 0.1$ & $29967.7 \pm 0.2$ & $0.1 \pm 0.2$ \\
NFW & 1 & 0.90 & $0.94 \pm 0.11$ & $1.01 \pm 0.12$ & $30033.1 \pm 0.1$ & $30033.2 \pm 0.1$ & $-0.1 \pm 0.2$ \\
$\alpha_{\rm Ein} = 0.2$ & 10 & 0.15 & $9.96 \pm 0.20$ & $10.06 \pm 0.19$ & $30088.8 \pm 0.2$ & $30088.8 \pm 0.2$ & $0.0 \pm 0.3$ \\
$\alpha_{\rm Ein} = 2.0$ & 10 & 0.15 & $10.01 \pm 0.17$ & $14.75 \pm 0.26$ & $30004.5 \pm 0.2$ & $30003.3 \pm 0.2$ & $1.2 \pm 0.3$ \\
$\alpha_{\rm Ein} = 0.05$ & 10 & 0.15 & $10.06 \pm 0.20$ & $9.48 \pm 0.19$ & $30060.5 \pm 0.2$ & $30060.1 \pm 0.2$ & $0.4 \pm 0.3$ \\
NFW & 10 & 0.15 & $9.44 \pm 0.19$ & $9.52 \pm 0.19$ & $29995.0 \pm 0.2$ & $29995.1 \pm 0.2$ & $-0.1 \pm 0.3$ \\
$\alpha_{\rm Ein} = 0.2$ & 10 & 0.90 & $10.00 \pm 0.31$ & $11.03 \pm 0.35$ & $30009.0 \pm 0.2$ & $30009.1 \pm 0.2$ & $-0.1 \pm 0.3$ \\
$\alpha_{\rm Ein} = 2.0$ & 10 & 0.90 & $9.85 \pm 0.27$ & $16.30 \pm 0.42$ & $30025.6 \pm 0.2$ & $30022.1 \pm 0.3$ & $3.5 \pm 0.4$ \\
$\alpha_{\rm Ein} = 0.05$ & 10 & 0.90 & $10.24 \pm 0.30$ & $10.93 \pm 0.34$ & $30208.4 \pm 0.2$ & $30208.4 \pm 0.2$ & $0.0 \pm 0.3$ \\
NFW & 10 & 0.90 & $8.82 \pm 0.31$ & $9.71 \pm 0.35$ & $29982.8 \pm 0.2$ & $29982.3 \pm 0.2$ & $0.5 \pm 0.3$ \\
\hline

\end{longtable}
\end{center}

\bsp	
\label{lastpage}
\end{landscape}
\restoregeometry
\end{document}